# Timekeeping at Akhet Khufu, as shown by the Diary of Merer

**Amelia Carolina Sparavigna**

Department of Applied Science and Technology, Polytechnic University of Turin, Turin, Italy

amelia.sparavigna @polito.it

*The discovery of the Diary of Merer (papyri Wadi al-Jarf) allows us to see the Egyptian calendar applied in a logbook. The diary is dated to the 26$^{th}$ year of reign of Khufu and describes Merer and his crew transporting the limestone blocks from the Tura quarries to Akhet Khufu, that is, the pyramid of Khufu (Old Kingdom). We find a calendar with 30-day months, 10-day weeks, five epagomenal days (Tallet, 2017) and Merer's job adapted to the seasons of the year. Seasons and epagomenal days have names according to inscriptions on the east wall of the Tehne tomb of Nj-k'-'nh (Nikaiankh), end of Fourth Dynasty or early Fifth Dynasty. The Diary gives us an overall impression of extreme modernity in the logistics of ancient Egypt. And Merer looks like a person with a reputation for good timekeeping at Akhet Khufu.*

## Introduction

In 2022, in an article entitled "Keeping time at Stonehenge", Timothy Darvill proposed that a solar calendar, with 30-day months, 10-days weeks plus five epagomenal days, was embodied into the erection of the Sarsen structures of the monument. "Recent remodeling of the developmental sequence at Stonehenge shows that the three sarsen structures—the Trilithons, Sarsen Circle and the Station Stone Rectangle—all belong to Stage 2 and were set up during the period 2620–2480 BC" (Darvill 2022, Darvill et al., 2012: Darvill, 2016). Therefore, the embodiment of a calendar happened around 2620-2480 BC. Darvill is also mentioning ancient Egypt: "External influences [at Stonehenge] may also be possible, … During the early third millennium BC, … increased interest in solar deities, such as the cult of Ra (Quirke, 2001), led to the development in Egypt of a 365-day solar calendar, known as the Civil Calendar. The origins, development and form of the Egyptian Civil Calendar have been described in detail by Nilsson (Nilsson, 1920) and Stern (Stern, 2012)" (Darvill, 2022).

In Darvill, 2022, we find a reference to a period when in Mesopotamia and Egypt we have calendars associated with written texts. Being calendars based on numerals it seems reasonable to consider that they existed before the advent of writing, even if we have no written accounts of them. For instance, people keeping time at Stonehenge could use calendars where solstices were relevant. Beyond any doubt Stonehenge was aligned with the summer solstice, that is, its main axis of symmetry is coincident with the sunrise on summer solstice (see the discussion in Darvill, 2022). "At Stonehenge, the emphasis on the solstices embedded into the architecture in the form of the principal axis and its orientation strongly suggests a solar-based system" (Darvill, 2022), for time reckoning.

Leaving aside the well-known solar orientations of monuments in Great Britain and Ireland, where the embedding of luni-solar calendars can be further investigated (see Darvill, 2022, and references therein), it seems that a rumored very old solar time reckoning existed. If we ask Google for the oldest calendar, the answer is that it appeared at Göbekli Tepe, made of some marks on a pillar (Göbekli Tepe is a 12000-year-old archaeological site in Turkey). In particular, August 2024 news told that the Göbekli Tepe calendar was a solar calendar. The calendar is given by decorations on a pillar, above the representation of a large vulture. Numerology is used to turn decorations into a calendar. However, according to Banning, 2023, a "plausible interpretation" of the iconography, that news claim being a calendar, is that it represents Göbekli Tepe's building complex, shown "with domed or vaulted roofs", or in a hybrid top-side-view with "an early form of perspective" (Banning, 2023). The Göbekli Tepe calendar is also the subject of a discussion proposed by Andre Costopoulos, where the related information flooding in the web is stressed. "This coverage has been almost universally uncritical. [Costopoulos] has seen no attempt to actually evaluate the claim, even though it should be obvious to even moderately informed readers that the study is highly speculative and does not support its findings adequately" (Costopoulos, 2024).

According to Darvill, 2023, we can define "numerology", as "the sense of constructing meaningful relationships between numerical patterns and concepts as a way of understanding the world". Some archaeoastronomers, that were irritated by Stonehenge "numerology", have not raised any concern about Göbekli Tepe calendar, which is based on numerology too. May be, Göbekli Tepe numerology is of archaeoastronomers' taste because based on astronomical interpretations of decorations, but the fact that no observation has been raised by them does not mean that the Göbekli Tepe calendar existed. The interpretation of the decoration is different, in particular of the disk above the vulture wing (see Collins, 2014, Banning, 2023).

Leaving the supposed Göbekli Tepe calendar aside too, let's investigate the Egyptian calendar. In this case we have written texts, then we can accompany the discussion with hieroglyphs. We will find the calendar used in the Merer's Diary, which is a logbook dated to the 26th year of reign of Khufu (Tallet, 2017). The Diary describes Merer and his crew transporting the limestone blocks from the Tura quarries to Akhet Khufu, that is, the pyramid of Khufu (Old Kingdom). Merer used a calendar with 30-day months, 10-day weeks and epagomenal days (Tallet, 2017, in the article on Wadi el-Jarf Papyrus H). Merer's job was adapted to the seasons of the year. Seasons and epagomenals have the names according to inscriptions on the east wall of the Tehne tomb of Nj-k'-'nh, early Fifth Dynasty, as shown by Der Manuelian, 1986. Thanks to the used calendar and the papyri "standardized layout", the Diary of Merer gives us an overall impression of extreme modernity in the logistics of ancient Egypt. And Merer looks like a person with a reputation for good timekeeping at Akhet Khufu.

Nj-k₃-ᶜnḫ   *On the left, the name that we find in Der Manuelian, 1986; here we write the name as Nj-k'-'nh*

**Civil calendar**

Darvill's words are stimulating a comparison of 'keeping time' in Egypt and its Civil Calendar of 30 days in a month. In fact, 30-day-months are coming from the time it takes the Moon to pass through all its phases (it takes about 29,5 days). In general, the lunar calendar contains 29-day and 30-day months. The Egyptian calendar had *fixed* 30-day months. As told by Spalinger, 2018, the fact "that the Egyptian civil months were based upon earlier lunar ones is easy to see"; moreover, "each civil month has a name as well as a number" (Spalinger, 2018). The Nj-k-'nh's table (early Fifth Dynasty) shows the case (Der Manuelian, 1986, name Nj-k-'nh is rendered as Nikaiankh by Thompson, 2014, and Friedman, 2015). The inscriptions in the tomb of Nikaiankh I, tell us of three seasons and months of each season numbered from I to IV. The name of the month is therefore an ordinal number. As we will show, the Merer's Diary of Fourth Dinasty contains the same manner of writing the months.

According to Spalinger, supposed irregularities of a previous lunar calendar led to the Civil Calendar, consisting of three "seasons" of four 30-day months. At the end of the year, 5 extra days (called "epagomenal", that is "added on") were added. About the link of the calendar with the Nile flood and the heliacal rising of Sothis (Sirius), see please the detailed discussion in Depuydt, 1997. Neugebauer considered the ancient Egyptian civil calendar "the only intelligent calendar which ever existed in human history" being based on fixed units - the 30-day-months - and simple to operate. He argued the Nile flood is the seasonal reason for the calendar.

For the history of the Civil Calendar, let us move to Egypt in the same period when the Sarsen structures at Stonehenge started to be developed. We find the building of the Great Pyramid of Giza, Akhet Khufu, the magnificent monument to pharaoh Khufu, of the Fourth Dynasty of the Old Kingdom. Akhet Khufu was built circa 2600 BC (Ramsey et al., 2010), over a period of about 26 years (Tallet, 2017). We know the name of the pyramid from ancient texts. Specifically, we know the name of Akhet Khufu from the [world's oldest papyrus](#) (in it Akhet is written with the ibis sign). The papyrus, one of the Red Sea papyri, was discovered at Wadi al-Jarf on the Red Sea coast of Egypt, 119 km south of Suez. The site hosted one of the oldest known artificial harbors in the world. Egyptian archaeologist Zahi Hawass called the Wadi-al-Jarf papyri "the greatest discovery in Egypt in the 21st century" ([bigthink](#)).

"Without a doubt, the Great Pyramid was commissioned by the Old Kingdom pharaoh Khufu (Cheops). The British Museum and Cairo's Egyptian Museum give his regnal dates as 2589 to 2566 BCE. Egyptologists Mark Lehner, who has conducted fieldwork at Giza for four decades, and Zahi Hawass, a former Egyptian government official in charge of Giza, argued for the later range of 2509 to 2483 BCE in their massive 2017 book, Giza and the Pyramids. But another high-profile Egyptologist, Pierre Tallet, whose pioneering fieldwork on the Red Sea coast of Egypt began in 2011, favors the earlier range of 2633 to 2605 BCE, derived from a recent astronomically based chronological model for the Old Kingdom" (Robinson, 2022).

**Red Sea Papyri**

"Astonishingly, the [Red Sea] papyri were written by men who participated in the building of the Great Pyramid, the tomb of the Pharaoh Khufu, the first and largest of the three colossal pyramids at Giza just outside modern Cairo. Among the papyri, it was the journal of a previously unknown official named Merer, who led a crew of some 200 men who traveled from one end of Egypt to the other picking up and delivering goods of one kind or another. Merer, who accounted for his time in half-day increments, mentions stopping at Tura, a town along the Nile famous for its limestone quarry, filling his boat with stone and taking it up the Nile River to Giza" (Stille, 2015). In literature we find different numbers of men in Merer's crew (the Elite). Let us therefore add that the base unit was of 40 men a phyle. "Phyles of the 4th Dynasty were in work crews of the pyramids building and the funerary architectural monuments, the Pyramid building was a national project combining workers and builders from different areas of Egypt" (M Abou Zeid et al., 2021).

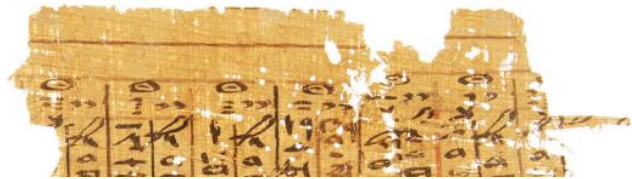

*Fig.1: Screenshot of the cover of Tallet's book. Many thanks to Pierre Tallet and AMeRS association for the document available at amers.hypotheses.org*

About the word Akhet (Khufu), rendered with the ibis sign, see please the cover of the book by Pierre Tallet, 2017, of the Diary of Merer. "The text, written with (hieratic) hieroglyphs, mostly consists of lists of the daily activities of Merer and his crew. The best-preserved sections (Papyrus Jarf A and B) document the transportation of white limestone blocks from the Tura quarries to Giza by boat" (Wikipedia and references therein). "Buried in front of man-made caves that served to store the boats at Wadi al-Jarf on the Red Sea coast, the papyri were found and excavated in 2013 by a French mission under the direction of archaeologists Pierre Tallet of Paris-Sorbonne University and Gregory Marouard. A popular account on the importance of this discovery was published by Pierre Tallet and Mark Lehner, calling the corpus "Red Sea scrolls" (an allusion to the Dead Sea Scrolls)." (Wikipedia and references therein).

"Though the [Merer's] diary does not specify where the stones were to be used or for what purpose, given the diary may date to what is widely considered the very end of Khufu's reign, Tallet believes they were most likely for cladding the outside of the Great Pyramid. About every ten days, two or three round trips were done, shipping perhaps 30 blocks of 2–3 tonnes each, amounting to 200 blocks per month. About forty boatmen worked under him. The period covered in the papyri extends from July to November" (Wikipedia and references therein). As we will see, also the cladding of the Bent Pyramid was made of limestone from Tura Caves.

Once more, see please the papyrus. In the image after restoration we have Khufu in six cartouches (article in The Past). Note the crested ibis (column seven from the left) of Akhet Khufu. Now, let us consider the upper part of the papyrus, and count from right to left. We can easily see the numbers 19, 20, 21, 22, 23, 24, and 25. These are the days of the month that we see in this part of the papyrus. How many days had the month? 30. Note that the days are separated by straight lines. Note please that the day is divided in two vertical columns, the right column is giving the diurnal activity, the left column indicates where Merer spent the night. Note the regular spacing: Merer had a ruler.

In the diary, we can see in the Sections BI and BII of Papyrus B, the end of a month and the beginning of the next month (Tallet, 2017) (see biblio.com)

"Section BI - … Day 28: casts off from Akhet-Khufu in the morning; sails *upriver* <towards> Tura South. Day 29: Inspector Merer spends the day with his phyle hauling stones in Tura South; spends the night at Tura South. Day 30: Inspector Merer spends the day with his phyle hauling stones in Tura South; spends the night at Tura South. Section B II - [First day <of the month>] the director of 6 Idjer[u] casts of for Heliopolis in a transport boat-iuat to bring us food from Heliopolis while the Elite (stp-s3) is in Tura. Day 2: Inspector Merer spends the day with his phyle hauling stones in Tura North; spends the night at Tura North. Day 3: Inspector Merer casts off from Tura North, sails towards Akhet-Khufu loaded with stone" (Tallet, 2017). In fact, Merer and his crew passed from Tura South to Tura North on the first day of the month (Tallet, 2017).

Thanks to the excavations made by the Tallet and Marouard teams we have the extraordinary possibility to observe 'modern' logistics in ancient Egypt, in the form of a report, directly from the hands of people that built the pyramids. We have the elegantly written papyri with the Egyptian Civil Calendar used to report the scheduled time of handling of stones. Merer's logbook then shows the accomplished logistic tasks in the scheduled time, with Merer looking like a person with a reputation for good timekeeping at Akhet Khufu.

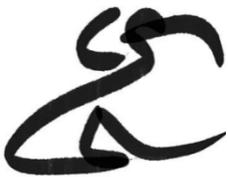
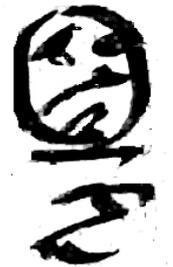

*Fig.2: On the left: Hieratic version of the crested ibis (adapted from the image courtesy of Wionvard, Wikipedia, CC BY-SA 4.0). On the right: See again the Merer's papyrus, day 22, second part of the day, below the Khufu's cartouche.*

**Merer's Diary**

"Having pieced together evidence written in both hieroglyphic and hieratic script, Tallet and Lehner expertly reveal the archives of a single work gang, 160 men in total, which cover slightly more than a calendar year, probably during the final year of Khufu's reign. The crew worked under the naval designation "The Escort Team of 'The Uraeus of Khufu is its Prow'" - the Uraeus being the stylized image of an upright, rearing cobra, a symbol of Egyptian sovereignty. Divided into four sections, they undertook at least five different tasks in different locations, all of which were recorded on separate papyri" (Robinson, 2022). Merer's crew was "The Elite".

Let us consider Papyrus B again.

[Day 25]: [Inspector Merer spends the day with his phyle [h]au[ling]? st[ones in Tura South]; spends the night at Tura South [Day 26]: Inspector Merer casts off with his phyle from Tura [South], loaded with stone, for Akhet-Khufu; spends the night at She-Khufu. Day 27: sets sail from She-Khufu, sails towards Akhet-Khufu, loaded with stone, spends the night at Akhet-Khufu. Day 28: casts off from Akhet-Khufu in the morning; sails upriver <towards> Tura South" (Tallet, 2017).

What is She-Khufu? "She Khufu" means "the pool of Khufu", and "Ro-She Khufu", the "entrance to the pool of Khufu", "which is perhaps the headquarters for the administration of the pyramid project, situated on the artificial lake near the mortuary temple" (Roger Pearse, roger-pearse.com). And Tura-South? The southern caves of Tura. And the Nile? Or other waterways?

"On a summer afternoon around 4,600 years ago, near the end of the reign of the pharaoh Khufu, a boat crewed by some 40 workers headed downstream on the *Nile* toward the Giza Plateau. The vessel …, was laden with large limestone blocks being transported from the Tura quarries on the eastern side of the Nile. Under the direction of their overseer, known as Inspector Merer, the team steered the boat west toward the plateau, passing through a gateway between a pair of raised mounds called the Ro-She Khufu, the Entrance to the Lake of Khufu. This lake was part of a network of artificial waterways and canals that had been dredged to allow boats to bring supplies right up to the plateau's edge" (Weiss, 2022).

"As the boatmen approached their docking station, they could see Khufu's Great Pyramid, called Akhet Khufu, or the Horizon of Khufu, soaring into the sky" (Weiss, 2022). Weiss is translating Akhet as 'horizon'. This translation has been questioned (Creighton, 2014), that is, the use of the term 'horizon' is misleading.

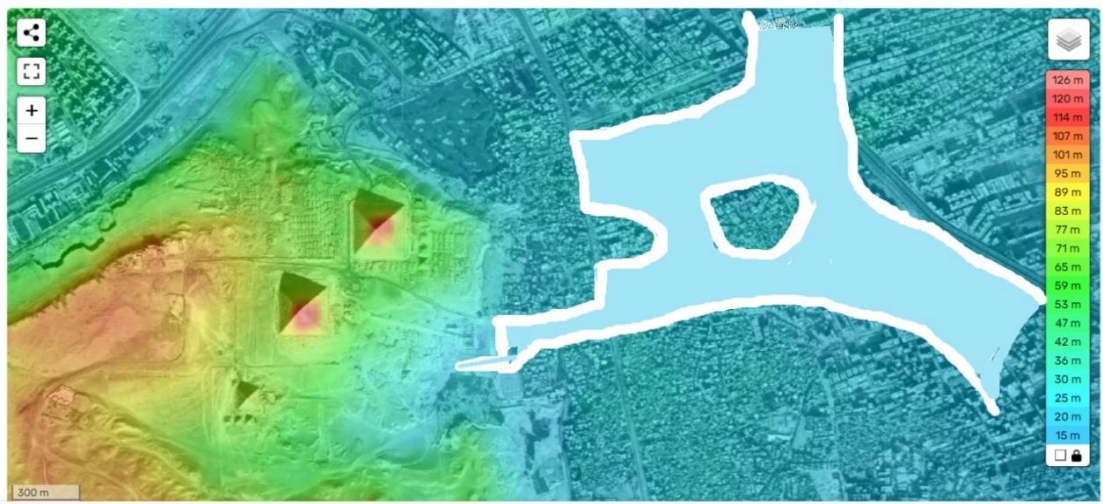

*Fig.3: On a TessaDEM elevation map (many thanks to the web site it-it.topographic-map.com), we have roughly sketched the waterway as given by Lehner, 2020, in his Fig.3. The caption of Lehner's figure tells that the map shows a reconstruction of 4th Dynasty water transport infrastructure at Giza.*

**Akhet Khufu's harbor**

In "Merer and the Sphinx", 2020, Mark Lehner illustrated his research about "geomorphological, archaeological, and architectural evidence related to the broad bedrock terrace on which the Khafre Valley Temple and Sphinx Temple were built to suggest that it may have served as a landing stage and depot for delivery of stone and other material needed for building the pyramid of Khufu". "The terrace would have been the easiest place for Merer and his phyle to offload the limestone that they delivered from Tura for the Khufu pyramid, as attested in the Wadi el-Jarf Papyri. A reconstruction of 4th Dynasty water transport infrastructure at Giza, based on evidence that has come to light over the past 30 years, suggests a central canal basin led from the Ro-She Khufu on a bank of a western Nile branch straight to the area east of the Sphinx. From here the shortest distance to the Khufu Pyramid leads along the course of the modern road passing 50 m north of the Sphinx. Does this reinforce the idea that Khufu conceived and started to form the Great Sphinx? While this is possible, architectural traces around the foundations of the Khafre Valley Temple and Sphinx Temple make it most probable that it was Khafre's workers who quarried and shaped the Sphinx and built the Sphinx Temple as part of the same quarry-construction sequence" (Lehner, 2020). The work by Lehner is full of detailed illustrations (see our Fig.3).

In Lehrer, 2020, we find also a comment about the old Nile channels. "The proximity and position of the nearest Nile channel was a major consideration for Merer and his deliveries by boat. Did he approach Giza via a secondary channel or the main trunk channel? A number of scholars who have looked at the question see an Old Kingdom Nile marked by the track of the very old Libeini Canal (Jeffreys and Tavares 1994; references collected in Lehner 2009, 102–110), which now flows with motor traffic, having been filled and turned into a four-lane asphalt highway. While today the Nile trunk channel splits into the Rosetta and Damietta branches north of Cairo, there is a reasonable chance that the bifurcation of the Nile took place south of Giza in the early Old Kingdom [see please references in Lehner, 2020], so *that two or more branches* could have flowed east of Giza at that time. [Lehner] modeled a Nile channel 500 m wide along the course of the Libeini with metrics like those of the Rosetta in the 19th to the early 20th centuries AD" (see Lehner, 2020, and references therein).

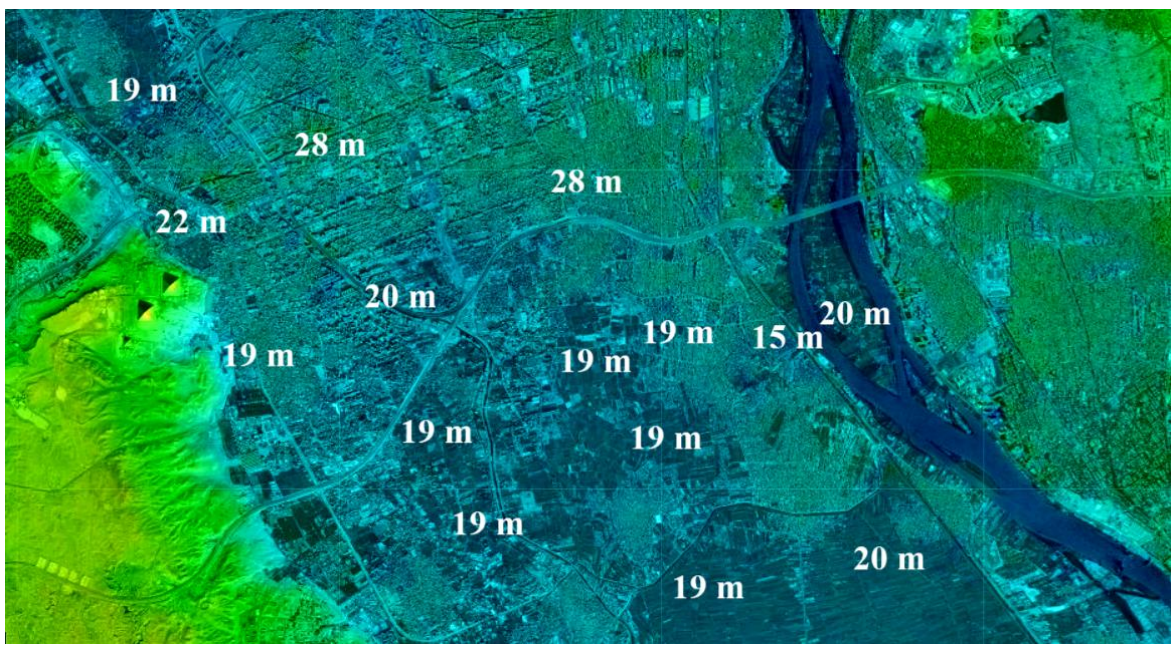

*Fig. 4: Thanks to [TessaDEM](), we can show the elevation of some locations between the Nile and the pyramids. How was the Nile at the time of Khufu? Recent research tells that the river was quite different than today.*

In Ghoneim et al., 2024, we can find a study where the researchers "identify segments of a major extinct Nile branch, which [they] name The Ahramat Branch, running at the foothills of the Western Desert Plateau, where the majority of the pyramids lie. Many of the pyramids, dating to the Old and Middle Kingdoms, have causeways that lead to the branch and terminate with Valley Temples which may have acted as river harbors along it in the past. [Ghoneim and coworkers] suggest that The Ahramat Branch played a role in the monuments' construction and that it was simultaneously active and used as a transportation waterway for workmen and building materials to the pyramids' sites" (Ghoneim et al., 2024). Ghoneim and coworkers are showing the Libeini waterway too.

In their Fig.7, Ghoneim et al., 2024, are showing a rendering of Giza landscape and its floodplain. Authors say in the caption that "TDX data shows, in 3D, a clear topographic expression of a segment of the former Ahramat Branch in the Nile floodplain in close proximity to the Giza Plateau". TDX means TanDEM-X radar.

"Besides the pyramid itself, the pyramid complex includes the mortuary temple next to the pyramid, a valley temple farther away from the pyramid on the edge of a waterbody, and a long sloping causeway that connects the two temples. A causeway is a ceremonial raised walkway, which provides access to the pyramid site and was part of the religious aspects of the pyramid itself (Ghoneim et al., mentioning Lehner, 1997). In the study area, it was found that many of the causeways of the pyramids run perpendicular to the course of the Ahramat Branch and terminate directly on its riverbank" (Ghoneim et al., 2024).

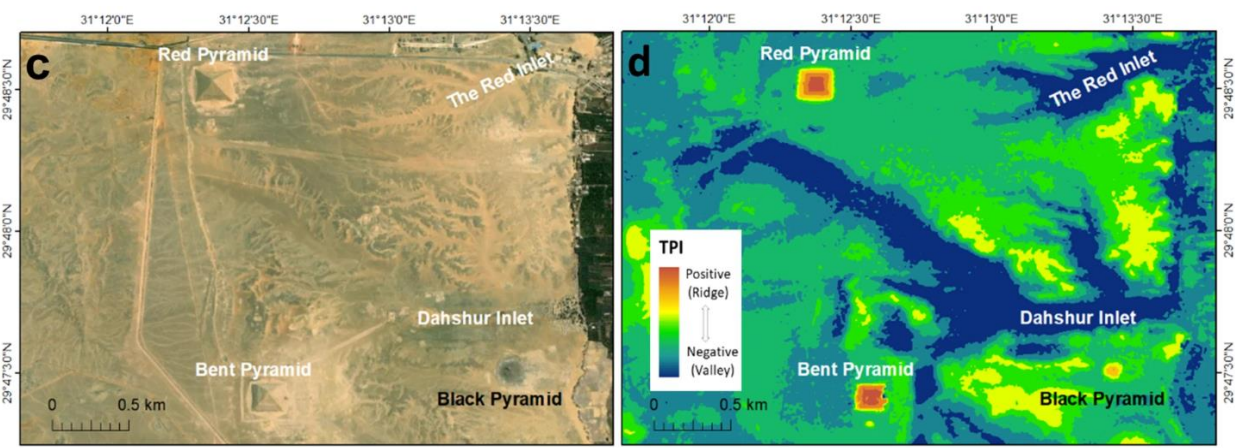

*Fig.5: Here it is reproduced a part of Fig.5 in Ghoneim et al., 2024 (many thanks to the authors who shared their article Open Access, CC BY 4.0).*

In Ghoneim et al., 2024, in their Figure 5, we can see in panels c and d the today "dry channels/inlets masked by desert sand in the Dahshur area". "The channels' courses were extracted using TPI. Negative TPI values highlight the courses of the channels while positive TPI signify their banks". TPI means Topographic Position Index. Let us remember that the Red and Bent Pyramids had been built by Old Kingdom Pharaoh Sneferu.

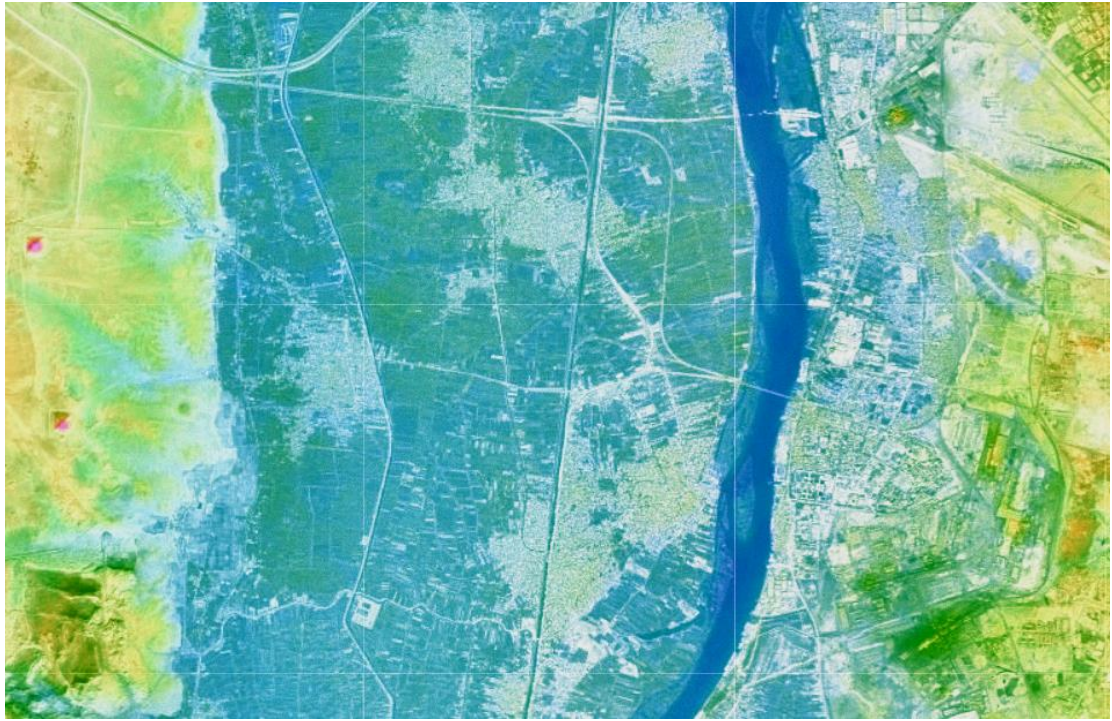

*Fig.6: A TessaDEM map, processed to enhance the presence of the inlets. The Red and Bent pyramids are on the left.*

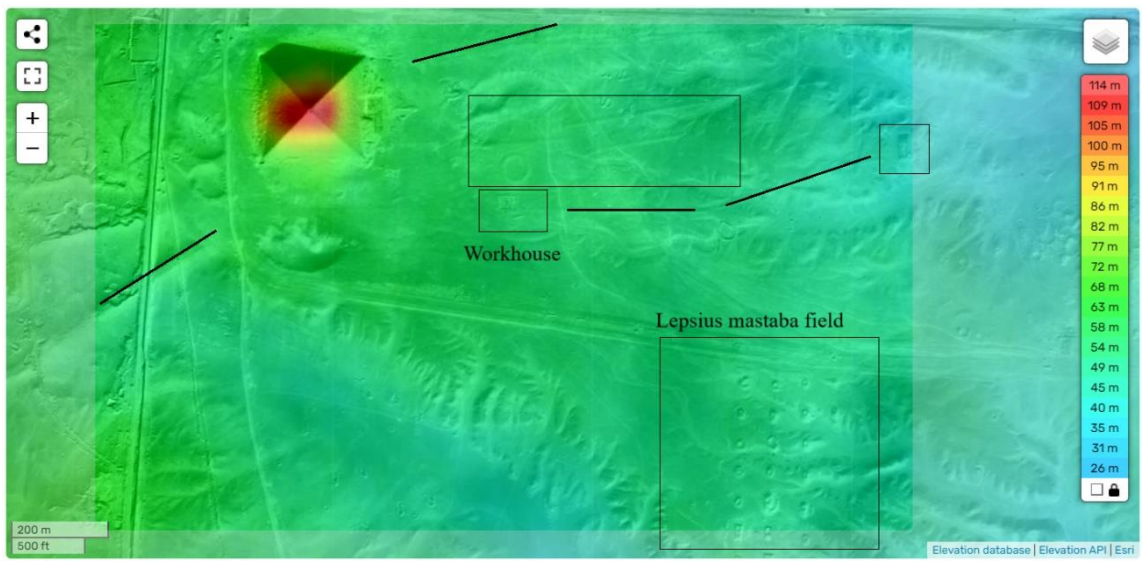

*Fig.7: The Red Pyramid is an image courtesy TessaDEM. Many thanks to TessaDEM for the remarkable maps here used for study and research. In the map we have marked some places we can find in the map by Moeller, 2016.*

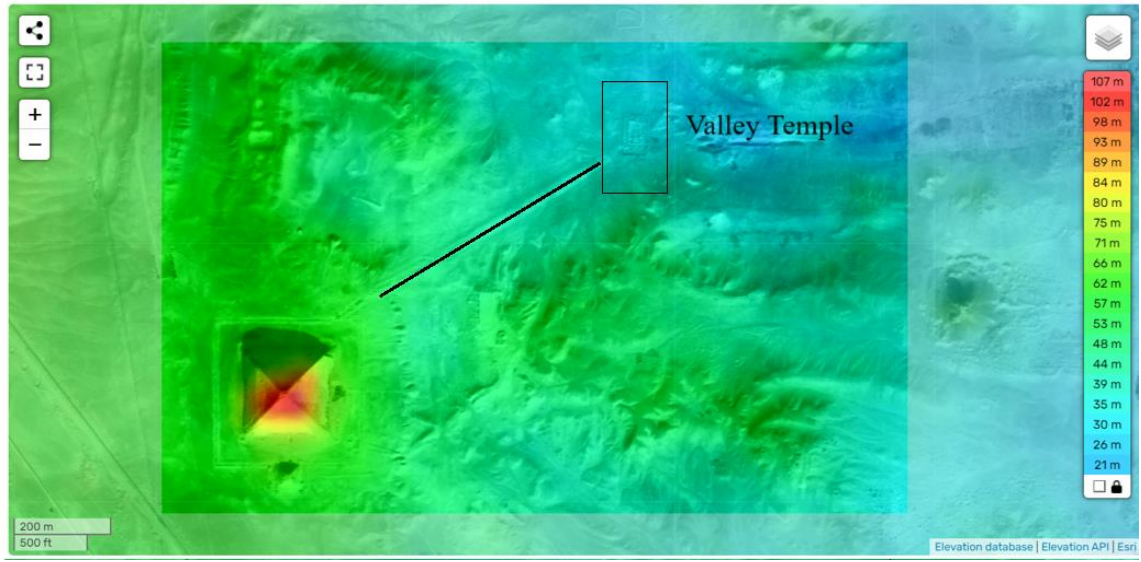

*Fig.8: The Bent Pyramid in an image courtesy TessaDEM. Note the causeway running north-east and the Valley Temple.*

In Fig.8, we used a TessaDEM elevation map to evidence the Bent pyramid, its causeway and the Valley Temple. According to Ghoneim et al., 2024, close to the Valley Temple there was the inlet of a today's dry channel.

The Bent Pyramid has a structure "made of limestone blocks, relatively thick, arranged in horizontal rows and covered with a layer of fine limestone from Tura" (Wikipedia, mentioning Maragioglio & Rinaldi, 1963). In MacKenzie et al., 2011, NMR spectra are provided of "the outer casing stone from Snefru's Bent Pyramid in Dahshour, Egypt, [compared] with two quarry limestones from the area. The NMR results suggest that the casing stones consist of limestone grains from the Tura quarry, cemented with an amorphous calcium-silicate gel formed by human intervention, by the addition of extra silica, possibly diatomaceous earth, from the Fayium area" (MacKenzie et al., 2011). A discussion about the quarries of limestone is given also in Tallet, 2017.

Since in the Introduction we have mentioned archaeoastronomy, let us observe that, besides the local environmental constraints for positioning pyramid, causeway and valley temple, we could propose astronomical orientations too. In Sparavigna, 2018, it has been shown that Bend Pyramid causeway has the same orientation of moonrise on a major lunar standstill and the rising of Capella. Then, why not consider the moon and the stars? Archaeoastronomy is structurally based on alignments according to sun, moon and stars, since its beginning with the studies of Heinrich Nissen and Norman Lockyer (as shown in a discussion of mine about Alexandria of Egypt, Nissen focused on alignments with stars of the main streets of this town). Moreover, archaeoastronomy is based on observations according to the horizon. Therefore, the Egyptian word "horizon" requires a detailed analysis.

**The Akhet environment**

Hong-Quan Zhang, 2023, has investigated the Akhet environment, as we can find depicted by the Pyramid Texts, comparing it with the local hydrology during the Green Sahara time. As Zhang tells in the abstract, "The Pyramid Texts contain vivid, specific, and consistent details of the Akhet and its surrounding lakes, canals, and farmlands." Using a climate change approach and hydrological profiles of the Green Sahara time, Zhang's "insight provides a key to decipher many obscure words in the current English translations. It also helps pinpoint the scattered environmental descriptions in the Pyramid Texts into a coherent portrait". The article "examines the relevant original hieroglyphs in the Pyramid Texts against their corresponding transliterations and English translations".

Zhang writes:

> The hieroglyph Akhet (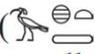 or 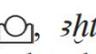, ȝḥt, meaning "place of becoming effective") is generally translated as "horizon" which blanks its original true identification.

Akhet Khufu is written, as previously said, with the ibis. This is clearly shown by the Wadi el-Jarf papyri (see our Fig.2). The other sign, Gardiner list N27, resembles the sun between two hills.

Robert Bauval, 2013, available Academia.edu, explained that, looking at sign N27 Mark Lehner in 1985 wrote: *[a] dramatic effect is created at sunset during the summer solstice as viewed, again, from the eastern niche of the Sphinx Temple. At this time, and from this vantage, the sun sets almost exactly midway between the Khufu and Khafre pyramids, thus construing the image of the Akhet, 'Horizon', hieroglyph on a scale of acres*". Mentioning archaeoastronomer G. Magli, Bauval tells that it has been concluded by Magli that Khufu and Khafre pyramids "were deliberately positioned relative to each other to create the hieroglyphic sign N27 'Akhet' i.e. "sun disk between two mountains", and this is why "Khufu called this project 'Akhet Khufu'" (Bauval, 2013). Bauval continued explaining that, "however, a *fatal flaw* [exists] with this idea: the hieroglyphic sign of the 'sun disk between two mountains' did not exist when Khufu built his pyramid! And even if it did, it was not used in the writing of the name 'Akhet Khufu'. In 1997 Lehner did acknowledge this fact: *Khufu's pyramid was Akhet Khufu. Here, and in the Pyramid Texts, Akhet is written with the crested-ibis and elliptical land-sign, not with the hieroglyph of the sun disk between two mountains that was used later to write 'horizon'.*" (Bauval, 2013).

Robert Bauval stresses that "sign N27 is not found <u>in the Pyramid Texts</u>, where the word 'Akhet' is written with the crested-ibis sign G25 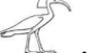, and the elliptical land-sign N18 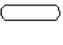. The same applies for the word 'Akhet' in the name 'Akhet Khufu', where 'Akhet' is also written with the crested-ibis sign G25 and elliptical land-sign N16. ... Amazingly, this mistake in interpretation of the name of the Great Pyramid is still found in recently published books. To be fair, Lehner [in his The Complete Pyramids, 1997, p. 29] did recognize that 'Akhet' - although erroneously translated as 'horizon' in many books - is written with the crested-ibis that also denotes the word Akh, meaning a 'spirit' who lives in the Duat (afterworld), the latter often written with a star in a circle". Bauval, 2013, continues adding that Lehner offered another meaning for 'Akhet Khufu', that is, *the place where the deceased (king Khufu) becomes an Akh, a suggested translation is "Spirit" or "Light Land"*. According to James P. Allen, who translated the Pyramid Texts, "The Axt (Akhet) is the place in which the king, like the sun and other celestial beings, undergoes the final transformation from the inertness of death and night to the form that allows him to live effectively - that is as an akh - in his new world. It is for this reason that the king and his celestial companions are said to "rise from the Axt (Akhet)," and not because the Axt (Akhet) is a place on the horizon or - as some have suggested - because it is a place of light.' Akhet, therefore, must mean the 'Place of Becoming Akh'..." (Bauval, 2013). This is the interpretation provided by J.P. Allen, as reported by Robert Bauval.

Please consider the discussion provided by Bauval. At the archived document, it is possible to see a picture of the supposed 'hierophany', the sun between two pyramids, first proposed by Mark Lehner in 1985, and reproposed by Magli. Bauval is discussing Magli and his arXiv, 2007, where we find Gardiner N27. In arXiv 2014, Magli added also the ibis-sign, but the so-called 'Khufu's project' of the two pyramids remains the same. That is, Magli's proposal is that Khufu planned two pyramids, that is that of Khufu and Khafre. Khafre, son of Khufu, built the second-largest pyramid at Giza. The name of the pyramid was Khafre-Wer which means "Khafre is Great" (Hawass, 2024).

For what is regarding the "dramatic effect", which "is created at sunset during the summer solstice", let us stress that the colossal statue of the Sphynx is facing east, and therefore not the sunset for sure. Please consider reading carefully the article by Lehner, 2020, entitled "Merer and the Sphinx".

Here is the conclusion of 2020 Lehner's work. "In our friendly dispute with our mentor, Rainer Stadelmann, Zahi [Hawass] once asked, "If Khufu made the Great Sphinx, what was its purpose?" This is a good question. Khufu's causeway extended some 800 m from his upper temple and swung far to the northeast to join his Valley Temple. The only possible reason for Khufu to carve the colossal Sphinx statue here would have been along the lines of the Colossus of Rhodes, or other harbor statues, the Statue of Liberty for an example in our time, in order to monumentalize the gateway to the Giza Necropolis, to the extent that city of the dead was under development flanking Khufu's Pyramid, as the pyramid itself rose (Jánosi 2005, 2006). These "elite" cemeteries were probably not accessible via the Khufu Valley Temple and causeway. Those courtiers who had been assigned the high-status mastabas must have ascended to the cemetery from the proposed landing place of Merer, at the end of Giza's central canal basin. But where else in Egypt do we find such a colossal image separated from a royal tomb or temple complex, save, perhaps, at Abu Simbel? True, like the Abu Simbel colossi, the Sphinx was part of a temple complex, and created as part of a quarry, construction, and image-shaping process. What is as certain as anything can be in archaeology, by virtue of the structural stratigraphy and by the Baugeschichte [building history] of the Sphinx, Khafre Valley Temple, and Sphinx Temple: It was Khafre's builders who carried out the greater part of that process and completed the Sphinx" (Lehner, 2020).

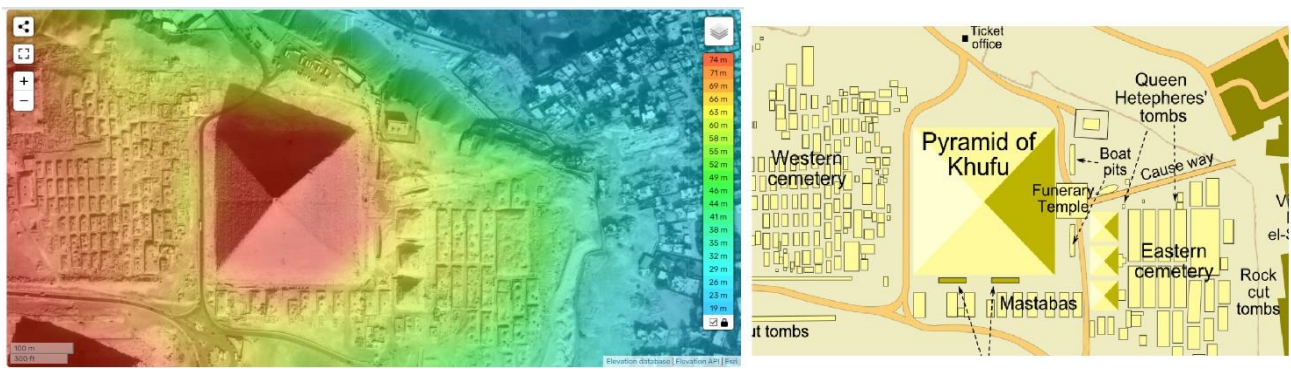

*Fig.9: Akhet Khufu in TessaDEM. On the right a detail oft he map courtesy MesserWoland, CC BY-SA 3.0.*

**Khufu and Contacts with the World**

Before returning to Merer's diary, let us consider Tallet, 2020. "Throughout the Old Kingdom period, the Egyptian state maintained close relations with all the regions surrounding the Nile valley. At the time when the pharaonic state launched monumental construction projects—notably the building of the gigantic pyramids of the Fourth Dynasty—the exploitation of mineral resources in the desert margins and in more distant areas was sharply accentuated. The establishment of harbors on the Red Sea shore served to reach the south of the Sinai Peninsula for the exploitation of copper and turquoise, as well as to bring back aromatics and exotic products from the land of Punt in the Bab el-Mandab area. The need for labor to realize building projects and develop the Egyptian infrastructure, for example as required to control major trade routes, led to repeated military raids against Libya, Nubia, and the Levant. Drawing on archaeology and written sources, including the tomb autobiographies of state officials of the Sixth Dynasty, [Pierre Tallet] offers perspectives on the complex military and diplomatic activities that linked the Old Kingdom to the surrounding regions" (Tallet, 2020).

Margaret Lucy Patterson, in a 2021 article, reported the Tallet's talk to the Essex Egyptology Group about Wadi el-Jarf. Tallet referred that "there are three Ancient Egyptian harbours known on the Red Sea coast of Egypt. As well as Wadi el Jarf there is another harbour to the north at Ayn Soukhna …, and one to the south called Mersa Gawasis which has been known since 1976. These harbours let us know how the Egyptians got to the Sinai and also to the still mysterious land of Punt. Tallet told [Essex Egyptology Group] that he would only be presenting the site of Wadi el Jarf in his talk. Sinai is visible from the Wadi el Jarf harbour site, and it's situated directly opposite a mining site in the Sinai which has evidence of Egyptian use dating back to Prehistoric Egypt. There were regular expeditions to the site from the Old Kingdom onward to mine copper for tools, as well as turquoise. *The Wadi el Jarf harbour was occupied in the early part of the 4th Dynasty and most evidence they've found dates to the reigns of Sneferu and Khufu*. It was abandoned after this, and the harbour that the expeditions used moved to Ayn Soukhna as the earliest evidence at that site dates to after Khufu's reign" (Patterson, 2021).

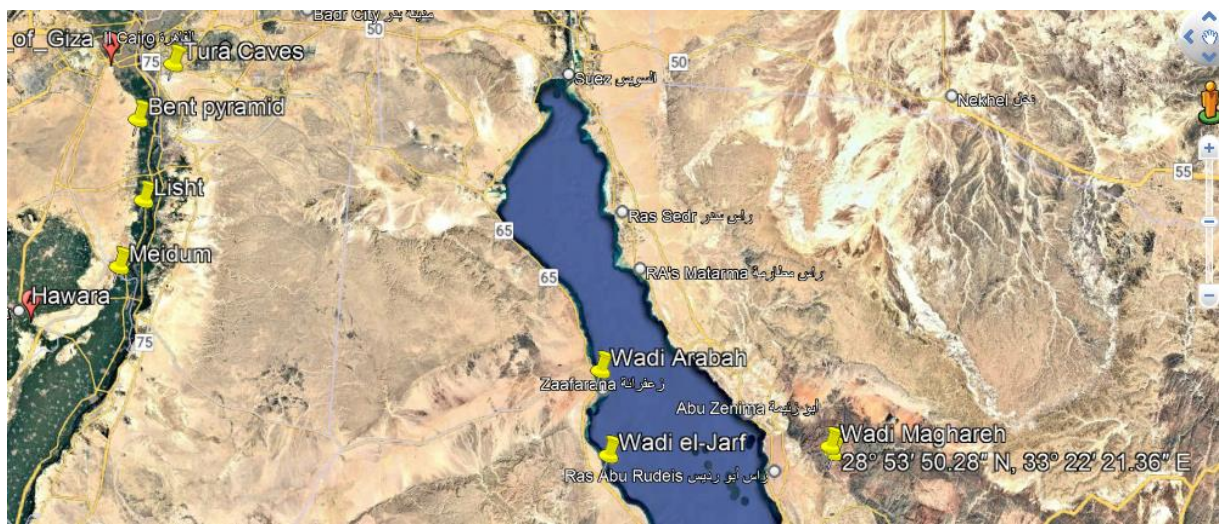

*Fig. 10: Map Courtesy Google Earth. See also biblio.com.*

"Tallet told [to the Essex Egyptology Group] that the [Egyptian] expeditions [at the time of Sneferu and Khufu] would have left Memphis and travelled by boat down to Meidum on the Nile. From there they would travel by land tracks to Wadi el Jarf through the Wadi Arabah [today, there is a modern asphalted road]. At Wadi el Jarf the expedition would travel by boat to el Markh – this site was excavated in the 20th Century and dates to the same time frame as Wadi el Jarf. The two harbours operate in tandem with boats shuttling back and forth between them, and they were only occupied and in use where there was an expedition in progress – there were cave systems at the Red Sea coast harbours to store the boats in when they were not in use. Once in Sinai the expedition travels by land to the Wadi Maghara (amongst other sites) where they mine copper and turquoise. There are reliefs at the mining sites from the time of Khufu, which have been known from the 19th Century onward" (Patterson, 2021).

**Merer's year**

Weiss, 2022, describes Merer's missions in the following manner.

"The [Merer's] logbooks cover the phyle's activities for just over a year near the very end of Khufu's tenure. In the summer months, roughly June to November, the workers operated in the vicinity of Giza. The beginning of this period was the first month of the inundation season, called Akhet, corresponding to the arrival of the annual Nile flood. The phyle's first assignment appears to have involved transporting around 600 workers to the Ro-She Khufu. There, Tallet believes, based on references in the logbooks to "works related to the dike of Ro-She Khufu," "the dam of the entrance of Khufu's lake," and the team "lifting the piles of the dike," these workers opened the floodgates to fill the basins and canals that allowed goods to be delivered to the foot of the pyramid complex construction site" (Weiss, 2022). "With these waterways open, Merer's men then spent several months hauling loads of fine limestone blocks from the Tura quarries to the Giza Plateau" (Weiss, 2022).

| Egyptian Month | Mission | Julian Month |
|---|---|---|
| Akhet I | Job at Ro-She-Khufu | July |
| Akhet II | Tura-Akhet-Khufu | August |
| Akhet III | Tura-Akhet-Khufu | September |
| Akhet IV | Tura-Akhet-Khufu | October |
| Peret I | Tura-Akhet-Khufu | November |
| Peret II | Nile Delta | December |
| Peret III | Nile Delta | January |
| Peret IV | Nile Valley | February |
| Shemu I | Nile Valley | March |
| Shemu II | Red Sea | April |
| Shemu III | Red Sea | May |
| Shemu IV | Red Sea | June |

"In December, when the Nile flood had ebbed and transporting heavy loads by boat to the Giza Plateau was no longer feasible", Merer and his crew were sent near two towns of the Nile Delta. The team was working at a facility structure for supporting "maritime expeditions to the Levant … 'At that time in history, the Egyptians were trying to connect as much as they could with the outer world,' says Tallet" (Weiss, 2022). "The papyri include no mention of the phyle's activities from January through March, and Tallet believes this may have been a season when the workers could return home to spend time with their families. Starting in April, the phyle appears to have been working at Wadi el-Jarf" (Weiss, 2022).

Merer used a calendar, with 30-day months and 10-day weeks; was it the Civil Calendar? Yes. Pierre Tallet indicated possible scheduled Merer's missions (see the Table here given). As we will discuss in the Section "Papyrus and Ink", there are also the Epagomenal Days.

Therefore, let us stress the importance of Merer's Diary for the history of Egyptian society, because it is showing the Civil Calendar use early in the reign of Khufu.

Wikipedia: "The civil calendar was established at some early date in or before the Old Kingdom, with probable evidence of its use early in the reign of Shepseskaf (c. 2510 BC, Dynasty IV) and certain attestation during the reign of Neferirkare (mid-25th century BC, Dynasty V)" (Wikipedia, 13 November 2024, is mentioning Clagett, 1995). "It was probably based upon astronomical observations of Sirius whose reappearance in the sky closely corresponded to the average onset of the Nile flood through the 5th and 4th millennium BC. A recent development is the discovery that the 30-day month of the Mesopotamian calendar dates as late as the Jemdet Nasr Period (late 4th-millennium BC), a time Egyptian culture was borrowing various objects and cultural features from the Fertile Crescent, leaving open the possibility that the main features of the calendar were borrowed in one direction or the other as well" (Wikipedia, 13 November 2024, Englund, 1988). "The closest approximation to the Mesopotamian system of the 3rd millennium might have been that of predynastic Egypt, where there seems to have been at once an administrative year of 12 30-day months with the addition of 5 days at the end of the year, parallel to which a synodical calendar remained in use" (Englund, 1988).

Shepseskaf is mentioned by Spalinger, 2018. Let us remember that Spalinger considers the lunar origin of the length of the months, that turned into the fixed 30-day months of the Civil Calendar. The list of kings is (Wikipedia): Sneferu, 2613–2589 BC, Khufu, 2589–2566 BC, Djedefre, 2566–2558 BC, Khafre, 2558–2532 BC, Bikheris, c. 2532 BC, Menkaure, 2532–2503 BC ?, Shepseskaf, 2503–2496 BC ?, Djedefptah (Existence disputed), 2496-2494 BC?. ***Then, the Merer's Diary is attesting the use of the Civil Calendar during the reign of Khufu (Tallet, 2017, Papyrus H), and therefore before the reign of Shepseskaf.***

**Merer's Pole Star**

"When the pyramid of Cheops was built around 2700 BC, the star Thuban, alpha Draconis, was the pole star. Every clear night in that epoch Thuban shone down a passage built into tha pyramid parallel to the polar axis" (O'Neil, 1976)

**Tura Caves**

"Merer gives two different itineraries for his round trips between the Tura quarries and Giza. Tallet found as a pattern in Merer's journal that when he and his team came from and went to Tura North they stopped overnight at Ro-she Khufu, the Entrance to the Basin of Khufu, but when they came from or returned to Tura South they stopped over in the She Khufu, Basin of Khufu, or Lake of Khufu" (Lehner, 2020). See please the discussion proposed by Lehner in his text of 2020.

The discussion is important because of the geographical location of the Tura Caves.

In Charlton, 1978, we find that "In 1940 it was decided to clear out the Tura Caves (note the word, indicating the view that they were a natural formation), with the object of storing the first portion of Middle East Forces reserves of ammunition. These caves were blocked with the accumulated bat dirt of thousands of years. … In spite of the name, Tura Caves were not a natural phenomenon, but a huge beautifully wrought quarry, or stone mine, whence had come the limestone blocks for the pyramids. There can be no disputing this, because in 1942 at one point at the end of the quarry *a pyramid block was found lying on wooden rollers*, still after some five thousand years awaiting delivery. On the side was a job code number in lamp black. To [Charlton] that job code number was and is the most moving relic of Pharaonic Egypt. It indicates the presence of a working engineer, part of that huge team of engineers and managers who first designed the programme for building a pyramid, and then carried it out. For those without experience, the design and layout of a construction job code itself demands an unusual combination of intelligence and industry" (Charlton, 1978).

"There are still problems. The normal explanation of the method of transporting the pyramid blocks from Ma'adi to Giza is that it was done by barge on the flood. But the distances are such that it is easy to think that the current at the peak of the flood would have carried the loaded barges well past Giza before they could have got across the river. And indeed, this must often have happened. The job must have been done on the waning flood, with the implication that the job of transporting the output of a year's quarrying and receiving the input for a year's construction must have been crammed into a few weeks. And that involves calculations of the numbers of men and barges employed which could indicate quite clearly that it was impossible to build the pyramids. But they were built, and they deserve the more their title of a wonder of the world the more closely their construction is scrutinized" (Charlton, 1978).

As we have seen from literature, canals and artificial waterways were used to control the navigation during the inundation season.

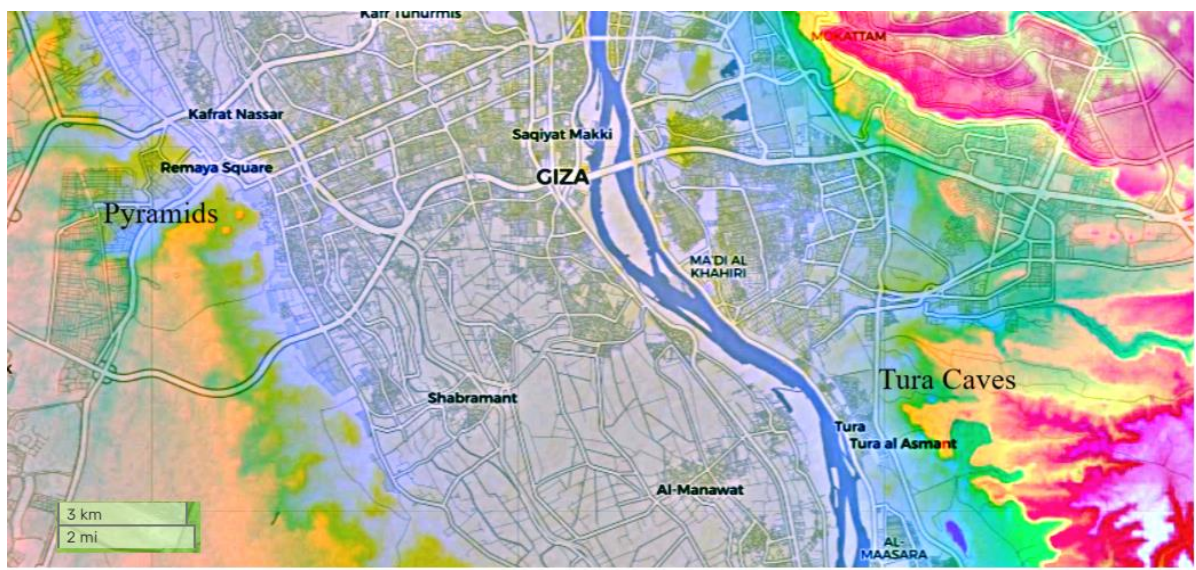

*Fig.11: Here we adapted a [TessaDEM](TessaDEM) map, showing the sites of the Giza Pyramids and the Tura Caves.*

In Figure 12, we can see the sites of Tura Caves and of the Giza Pyramids. At Khufu time there was an Eastern branch, such as a Western dry branch of the Nile (Lehner is mentioning two or more branches). Therefore, there was some waterway that Merer used to pass from the Eastern to the Western branch.

"One complicated feature of river travel in ancient times was that the course of the Nile shifted, so that the route from Tura to Giza most certainly would *not be as direct as it is today*. It has been suggested that Khufu and those before him constructed waterways and canals that connected the pyramid-building to harbors, reservoirs, and basins that would ease the passage of stone barges to the site" (Gaudet, 2019). In fact, Merer's crew worked at Ro-She Khufu too, the "entrance to the pool of Khufu".

In Papyrus A ([Merer's Diary](#)) we have: "Day 8: Cast off in the morning from Tura, sail downriver towards Akhet-Khufu, spend the night there. Day 9: Cast off in the morning from Akhet-Khufu, sail upriver; spend the night. Day 10: Cast off from Tura, moor in Akhet-Khufu".

It is clear: downriver and upriver.

As told by Lehner, 2020, Tallet has proposed two different waterways that Merer and his crew used to move from Tura North and from Tura South. So, let us consider [Pierre Tallet](#) in detail, "In the wider view", the Merer's papyri allow us to investigate "the Memphite region at a very ancient period in its history. This was a strategic zone for the Egyptian state, which in this period maintained its administrative capital here, just down from the ancient necropolis of Saqqara. Certain reconstitutions of the former course of the river can explain this choice by the split in the single course of the Nile precisely at this point, dividing into two branches much further south than today. The western branch, much further west than today, follows more or less the line of the present Bahr el-Libeini, a little more than a kilometre to the east of the Giza plateau" (Tallet, 2017). "Merer's Logbook does not provide confirmation of this geological phenomenon, but it does nevertheless present a rather novel image of the area between Tura and Giza, a zone marked by *large hydraulic developments maintained by the State* and where opposite points on the major course of the river were directly in contact solely by water transport" (Tallet, 2017). Ro-She Khufu, the gateway to the pool of Khufu, "should probably be sought in *one of the openings into the artificial lake* dug at the foot of the Giza necropolis, and within a radius of some 10 km maximum from the funerary complex of Cheops. As this point is only a stage for when the boats come from the most northerly quarries—Tura North—one might imagine an entrance to the ensemble situated relatively to the east of this great basin that lay at the foot of the Giza plateau. According to information provided by papyrus A, we may be dealing with *a strategic point, which, at the time of the flood, allowed control* of the workings of the overall hydraulic system that provided access to the Giza complex, through the lifting of a barrage installed at the entrance to a canal". (Tallet, 2017). See please the discussion provided by Tallet, in Annexe I, translated from French by Colin Clement.

**From control marks to logbooks**

The logbooks allow us to have information about the organization of the work of phyles under Khufu. "Those involved in the construction of royal monuments were until now simply known by the *control marks* that they left on the different sites of the royal funeral complexes" (Tallet, 2017). Thanks to the logbooks, we can determine a division into four squads, that Tallet estimates comprising of 40 men, under the leadership of an inspector. "Administrative control at the level of overall team was most likely in the hands of a scribe … Most probably, at the head of the chain command, with general authority over the labor unites, stood the project manager of the pyramid construction site, who, in the era of these archives, was the vizier, director of Ro-She Khufu, in charge of the royal works," half-brother of Khufu (Tallet, 2017).

It is specified in the Papyrus B that people of the crew "are required to accomplish a particular task, and it would seem that in Papyrus D, which most probably corresponds to the 1st month of Akhet at the very beginning of the calendar year, there is mention of an official document, perhaps a royal decree, which could be an official accreditation of the team for the coming 12 months. … Papyrus D shows that the team was familiar with the seats of the administration and power" (Tallet, 2017)

**Papyrus and Ink**

"Further major change came with the invention of papyrus as a writing and recording medium in the early years of the third millennium. Papyrus offered a larger writing surface than bone or ivory tags and allowed records to be made at greater length; the use of tags seems to have ended when papyrus became widely available. Clay sealing, however, continued to be used … . By the middle of the third millennium, detailed accounting records on papyrus were well-established in Egypt and allowed precise monitoring of revenues and expenditures. Surviving examples include accounts of food delivered to a team of workers [Tallet and Marouard, 2014] and accounts of grain and cloth delivered to the administrator of an agricultural estate [Posener-Kriéger, 1986]. Like their Mesopotamian counterparts, the earliest Egyptian accounting records did not attempt to encode all the complexities of the spoken language, but they were often partitioned into segments and made us of tabular formats to underpin their comprehensibility [Eyre, 2013]" (Yeo, 2021, and reference therein).

According to Yeo, Egyptians "took full advantage" of the new writing device. The papyrus allowed "to accommodate lengthy documentation"; in fact, we have surviving fragments but also scrolls the length of which is varying from 1 ¼ to 1 ½ meters (Yeo, 2021, mentioning Posener-Kriéger, 1986).

Yeo continues accounting for the Egyptian writing capacities on papyrus substrates.

We have mentioned before Margaret Lucy Patterson and her 2021 article, reporting the Tallet's talk to the Essex Egyptology Group about Wadi el-Jarf. Tallet referred that at Wadi el-Jarf there is "a large building situated between the camps and the harbour. ... At the time of Khufu this was a large rectangular building measuring about 40m by 60m, subdivided into long thin rooms (each room running the full length of the short dimension of the building). Each of these rooms could shelter 50 men, and it looks a lot like some of the buildings excavated near the pyramids at Giza". The building was "cleared out by its last Ancient Egyptian occupants when they packed up and left" the site and this fact "was rather disappointing" for Tallet and his team. "But underneath this building they have found another level of occupation – and this still had a lot of evidence of its occupation, which was quite exciting! This earlier building dates to the time of Sneferu (the predecessor of Khufu). They have found *many ostraca* dating to the time of Sneferu with inscriptions" (Patterson, 2021). Some ostraca were referring to the "heqat" unit of measurement. "As well as these there are also other ostraca that have the names of work teams, or the names of officials with their titles. … Another recent find at this part of the site is a shell (from the Nile Valley rather than local) which contains traces of *both red and black ink* on opposite sides of the shell. Tallet thinks this is probably part of a scribe's toolkit – his ink well" (Patterson, 2021).

Tallet continued describing the boathouses. Then he arrived to describe the papyri. Thanks to Patterson we have further information. The documents are dating to the "year after the 13th census of the cattle of Khufu", as Patterson is referring to us. "This would be year 26 or 27 of his reign (the cattle census was mostly done every 2 years, tho[ugh] this is not always the case), which is pretty close to the end of Khufu's reign (but no longer thought to be right at the end)" (Patterson, 2021). The workers' team had the name of "Khufu's uraeus is its prow", "and this team designation also appears on lots of locally produced storage jars. The team is run by an Inspector called Merer, and the documents also frequently name a scribe called Dedi. There are 6 papyri which cover the whole of a single year's work done by this team … . Each section has the month as the top row of the document, and the documents each cover 2 months of the year. Below the month is a row with all the days of the month in order, and below each day heading is two columns of hieratic text detailing the team's activities that day. Each week of 10 days (sometimes referred to as a decade, which [Patterson] find a bit confusing) is separated from the next by a red line" (Patterson, 2021).

Please consider again the picture of the Merer's papyrus. Note the *red* line between the day 20 and the day 21.

From the fragments of another document, thanks to the papyri "standardized layout", Tallet was able to investigate "the reciprocal nature of the relationship between the king and the workers – the documents has some notations in *black* ink which talk about what the workers are doing, and then others in *red* ink which show what they are receiving from the state in return (like clothing or equipment)... these documents show that the workforce were not slaves. The document also demonstrates that this is a multi-skilled team – they are not just stone transporters, they are also involved in the cult of the king during his lifetime" (Patterson, 2021).

About the calendar, we can add further information from Tiziana Giuliani, 2018. We can see a picture showing a "Frammento in cui viene indicato precisamente l'anno in cui sono stati redatti questi registri". A piece of papyrus where we clearly find the year during which the registers were written. From the "papiro del registro del pane, individuabile anche dai segni geroglifici posti appositamente sul papiro una volta arrotolato, troviamo un elenco con informazioni relative al mese in cui è stato redatto il documento, il luogo di provenienza del cibo, l'elenco dei viveri e le quantità di farina di grano, farina di orzo, grano e la data di consegna del pane che proveniva dal nomo dell'Arpione nel Delta Occidentale nei pressi dell'attuale città di Rosetta" (Giuliani, 2018). From the papyrus of bread register (Wadi ef-Jarf Papyrus H), which is also identified by the hieroglyphic signs specifically written on the papyrus once rolled, we find a list with information relating to the month in which the document was written, the place of origin of the food, the list of provisions and the quantities of wheat flour, barley flour, wheat and the date of delivery of the bread that came from the "nomo" of the Harpoon in the Western Delta near the current city of Rosetta.

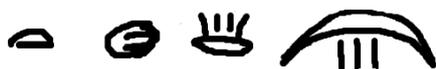

The sketch given on the left shows the name of a month written in the Wadi el-Jarf papyrus (sketch adapted from an image by Giuliani, 2018). It is the third month of the Inundation season.

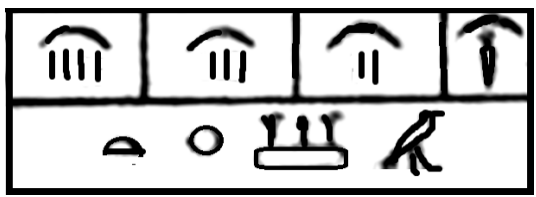

We provide the name of month and season according to Der Manuelian, 1986, sketch given on the left, thanks to an Old Kingdom relief sculpture on the east wall of the Tehne tomb of Nj-k'-'nh (see the following discussion).

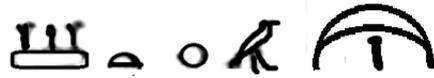

In the Papyrus A of Wadi el-Jarf, we have first day of the first month of the inundation, as we can see at the web site Académie des Inscriptions (Figure 3).

Section AI (Tallet, 2017): "First day <of the month>: […] spend the day […] in […]. [Day] 2: […] spend the day […] in? […]. [Day 3: Cast off from?] the royal palace? [… sail]ing [upriver] towards Tura, spend the night there. Day [4]: Cast off from Tura, morning sail downriver towards Akhet-Khufu, spend the night. [Day] 5: Cast off from Tura in the afternoon, sail towards Akhet-Khufu. Day 6: Cast off from Akhet-Khufu and sail upriver towards Tura […]. [Jour 7]: Cast off in the morning from […] Day 8: Cast off in the morning from Tura, sail downriver towards Akhet-Khufu, spend the night there. Day 9: Cast off in the morning from Akhet-Khufu, sail upriver; spend the night. Day 10: Cast off from Tura, moor in Akhet-Khufu" (Tallet, 2017). In the image, we can see the ibis sign for Akhet.

Let us consider the image (archived) provided by Tiziana Giuliani of the papyrus of bread register (Papyrus H). Also this picture, from the web site Meretseger Books, allows us to see the papyrus. Using the two images let us propose the following sketch:

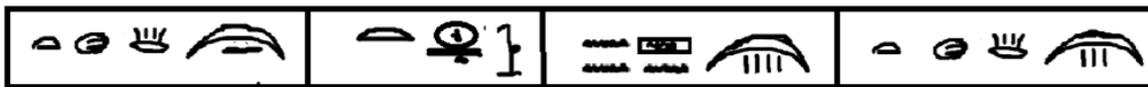

From right to left, we see the third month of the Inundation, then the fourth month of Shemu, then some written text, and the first month of inundation (horizontal line below the crescent). What is the meaning of the written text? In Steele, 2007, it is told that the earliest representation of the division of year into three seasons plus five epagomenal days is from about Mycerinos time, in an inscription from the tomb of Nj-k'-'nh at Tehne. Let us consider Der Manuelian, 1986, again. See please the Figure 3 in his article (caption: Detail sketch of duty table of Nj-k3-'nh showing the position of hrjw rnpt).

Please consider the screenshot of Der Manuelian's text:

> Although the writing of the epagomenal days does vary, for example, ⊙⦙⦙{𓀀𓏥} (Neuserre)[13] and {𓀀𓏥}⊙⦙⦙ (Pyr. §1961C, Pepi II), one always expects the "five days"⊙⦙⦙) to be written, or else the meaning of the phrase itself *ḥrjw rnpt* "which is on the year" would make little sense; but the ⊙⦙⦙ are missing from *Nj-k3-ʿnḫ*'s two tables.

Reference 13 in Der Manuelian is Kees, 1928. In the Wadi el-Jarf papyrus the signs of the five days are missing. But in the Wadi el-Jarf papyrus we are dealing with a calendar for sure. Therefore, we have 30-day months, 10-day weeks, plus epagomenal days. We have the Civil Calendar. If we consider the line under the crescent as indicating the first month, then the epagomenal time is between the last month of the year and the first month of the following year.

**Pierre Tallet and the Papyrus H**

After our exercise in reading the manuscript, where we helped ourselves with Der Manuelian, 1986, let us verify the results with Pierre Tallet.

"Le papyrus H a été découvert en deux gros fragments le 1er avril 2013. … La partie de droite du papyrus, d'une hauteur max. de 22 cm et d'une longueur de 45 cm, était la mieux préservée ; au recto, …. au verso est encore inscrit le titre de la pièce comptable, de façon à ce qu'il apparaisse sur l'extérieur du papyrus lorsque celui-ci était roulé. … 'compte de pain'"(Tallet, 2017).

Let us pass to the comments provided by Tallet. "La première date livrée par le document est le 3$^e$ mois de la saison Akhet, ce qui est très surprenant, puisque la section suivante enregistre le 4$^e$ mois de Chémou. Cela suppose en effet qu'il y a à cet endroit un « vide » de 8 mois dans la consignation des informations. Les sections mensuelles s'enchaînent ensuite logiquement, puisqu'apparaissent successivement le 4$^e$ mois de la saison Chémou, les jours épagomènes de la fin de l'année, et le 1er mois de Akhet. Deux solutions peuvent expliquer cette disposition inattendue des informations. …" (Tallet, 2017). The first date given by the document is the 3rd month of the Akhet season, then we find the 4th month of Chemou. Apparently, we have a "gap" of 8 months in the recording of information. The monthly sections then follow consequently and after the 4th month of the Shemu season we have the epagomenal days of the end of the year, and the 1st month of Akhet again. [Good, the horizontal line under the crescent represents the first month]. The arrangement of the first two monthly sections of the document seems closely linked, which probably means that they were written at the same time: the deliveries that are mentioned come in both cases from the Harpon nome, whose capital name was written astride the line separating the two divisions, as explained by Tallet.

Let us pass to the comment about the epagomenal days. "ḥrjw rnpt [are] les cinq jours supplémentaires qui s'ajoutent à l'année civile pour faire le compte de 365 apparaissent *à notre connaissance pour la première fois dans l'histoire égyptienne* sur ce document – la première attestation qui en était jusqu'ici généralement acceptée étant le calendrier d'offrande du temple solaire Niouserrê. Ils bénéficient d'une section de la comptabilité indépendante, qui enregistre un compte de denrée exactement proportionnel à cette durée, soit 1/6 de ce qui est généralement livré pour un mois de 30 jours. On note toutefois que le scribe n'a pas aussi clairement isolé cette section de la comptabilité que les autres". (Tallet, 2017). The epagomenal days have an independent accounting section, which records a supplied quantity exactly proportional to the time duration, i.e. 1/6 of what is generally delivered for a month of 30 days. It is noted, however, that the scribe has not as clearly isolated this section of accounting as the others. Since the quantity is 1/6 of the monthly supply, it means that the days are five. ***To Tallet's best knowledge, in this Wadi el-Jarf papyrus. we find the first attestation of the epagomenal days***.

Pierre Tallet notes that « Ce document maintient donc une certaine ambiguïté sur l'appartenance réelle de ces cinq jours, dont on s'est longtemps demandé s'ils étaient considérés comme les derniers jours de l'année finissante, ou au contraire comme les premiers jours de la nouvelle année ». The Wadi el-Jarf document maintains a certain ambiguity about the real position of these five days, of which it has long been wondered whether they were considered as the last days of the year, or on the contrary as the first days of the new year.

In a note it is also added that, «l'identification de ces jours supplémentaires aux premiers jours de l'année pourrait être suggérée par les listes d'offrandes de la chapelle de Nikaiankh Ier à Tehna – sans doute à dater de la fin de la IVe dynastie – où cette notation apparaît à deux reprises devant la mention de la saison Akhet» (Tallet, 2017). We will discuss in the next section.

**Seasons and epagomenals**

From Steele, 2007. "The Egyptians were the first to use the number 365 for calendrical purposes, … The earlies secure attestation of 365 as the number of the days of the Egyptian year seems to date to the reign of Mycerinus [Menkaure]. In the annals on the Palermo-stone, the number 365 can be inferred from how the transition from Mycerinus to Shepsekaf is represented. Erst beim Regierungswechsel von Mykerinus zu Schepses-kaf - writes von Beckerath - lässt sich der Gebrauch des Kalenders von 365 Tagen nachweisen, wie Borchardt demonstriert hat. In fact, the earliest comprehensive representation of the divisions of the year into seasons and epagomenal days also dates around the time of Mycerinos. The representation is found in an inscription from the tomb of N.k-'nh at Tehne. Mycerinus is mentioned in the inscription. The epagomenal days are listed in the table before the three seasons." (Steele, 2007). Nj-k'-'nh is Tallet's Nikaiankh.

The Merer's diary was discovered after 2007.

From Der Manuelian, 1986. "Among the unique and unparalleled specimens of Old Kingdom relief sculpture is the important but misunderstood "duty table" on the east wall of the second, or newer, Tehne tomb of Nj-k'-'nh (= tomb no. 1), … during the early Fifth Dynasty."

"Two important features of Nj-k'-'nh's table have, in my [Der Manuelian] opinion, been misinterpreted. The first is the caption 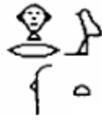 ; the argument has enjoyed wide acceptance that this is not the earliest attestation of hrjw rnpt, the so-called five epagomenal days. The second is the extent to which the table on Nj-k'-'nh's tomb wall accurately represents a stone version of an original papyrus document. Recent opinion has it that the table is a composite scene containing excerpts from numerous decrees and other documents. An attempt is made below to demonstrate that Nj-k'-'nh's table does indeed contain the earliest known reference to the epagomenal days and could well represent a copy of a single, original legal document in its entirety. Both problems revolve around the hrjw rnpt caption, and I [Der Manuelian] will …" (Der Manuelian, 1986). After a detailed discussion, Der Manuelian concluded "If such is the case, Nj-k'-'nh's table would serve as our earliest example, just as it gives us our earliest attestation of hrjw rnpt, the five epagomenal days".

The table in the tomb has a horizontal row where we see the name of the "Keeper of the King's Property and Steward of the Great Domain Nj-k'-'nh; his wife, the Keeper of the King's Property Hdt-hknw, and her children." "The next horizontal row bears the separately listed names of Hdt-hknw and her children along with some additional ka-priests. The "determinatives" or large-scale representations of the individuals fill up the next row of boxes, while underneath them appear first the individual months of the year, one per person, then the *three seasons indicated immediately* below them. … Below the two lines occupied by text C begins the table for the nobleman Hnw-k. Once again *come the individual months* and below them the *three season names*. Still further below these we find the list of children again, but this time without the large "determinative" figures and in reverse order. Thus, the timetable remains constant to both cults, but the personnel involved shifts positions" (Der Manuelian, 1986). See please the discussion of the epagomenal days proposed by Manuelian. We find mentioned Neuserre. "The entry on "Epagomenen" in the Lexikon der Agyptologie follows suit, citing E. Winter and dating the first occurrence of the epagomenal days to the offering list of Neuserre" (Der Manuelian, 1986).

In Der Manuelian we find discussed the position of the five days. The author is also giving a "brief review of selected publications" about the "disputed meaning of hrjw rnpt. Fraser labeled it "new year's day" in his translation. Both G. Maspero and J. H. Breasted preferred to read the "five epagomenal days", … In 1960, however, E. Winter proposed an alternative interpretation for hrjw rnpt, prompted by what he saw as four serious problems connected to the original epagomenal days explanation … E. Edel accepted Winter's explanation and … P. Posener-Krieger followed suit in her translation and commentary to the Abu Sir papyri. Mrsich made no reference to Winter in his monograph on Egyptian "house-documents," preferring to reconstruct a thirteen-month year, since there are thirteen columns or boxes (one of which was left blank) in the table. H. Goedicke not only agreed with Winter but cited … Helck both follows Winter and expands his interpretation of Nj-k -'nh's table accordingly to argue for copyist errors in the transmission from papyrus to tomb wall" (Der Manuelian, 1986, and references therein).

In [Figure 1 by Der Manuelian, 1986](), it is easy to find the three seasons and the four months in them (characterized by numbers). Note the last season, there is an empty box.

The "scholarly opinion rests at this point in substantial agreement with Winter, that Nj-k -'nh's hrjw rnpt is not our earliest reference to the five epagomenal days but a simple caption heading, like 'ht, "fields," designating the monthly divisions of the Egyptian calendar ("what is on/in the year"). The entry on "Epagomenen" in the Lexikon der Agyptologie follows suit, citing Winter and dating the first occurrence of the epagomenal days to the offering list of Neuserre" (Der Manuelian, 1986, and references therein). As told before, Der Manuelian concluded that "Nj-k'-'nh's table would serve as our earliest example, just as it gives us our earliest attestation of hrjw rnpt, the five epagomenal days". Today, we have the Merer's diary. Then, at the Khufu's time, had the calendar the epagomenals? Yes.

In the papyrus there are months and Epagomenals. Then, the Merer's calendar has 360 days + 5 days.

"It was after the final day of the twelfth month, Mesore (4 Smw), that the agents of the goddess Sekhmet were sent not merely to harm but also virtually to annihilate man. As early as the Pyramid texts we read of them as days of the "Birth of the Gods," a tradition that is first specified in one of the Harhotpe documents of the Early Middle Kingdom. At that time, all of the five deities associated with each day are recorded: Osiris, Horus, Seth, Isis, and Nephthys. … References to the epagomenals, of course, can be found in such texts as the Niuserre Temple Calendar of Dynasty V or the private feast lists of Khnumhotep 2 at Beni Hasan and those recorded on the Coffin of Ma, the latter two dated to the Middle Kingdom, as well as one of the Harhotpe documents as previously noted. But perhaps of greater significance to a chronologist than to a student of Egyptian religion *is the frequent occurrence of a truncated or simplified civil year comprising only 360 days*, a situation which has been recently presented by C. Leitz. The "absence" of the five epagomenals was part and parcel of the water clocks known in the Nile valley as early as the close of Dynasty XVIII. Other evidence, also covered by Leitz, gives additional support to the feeling that these five days need not have been placed in all liturgical or calendrical systems" (Spalinger, 1995).

"The 365-day calendar was itself a modified version of an earlier calendar of 360-day years. Evidently, at some point in the second half of the third millennium BC, the Egyptians added in five additional ("epagomenal") days to the 360-day year, but apparently realized that this did not work either. Thus, amazing enough, in economic, administrative and sometimes even astronomical texts, the Egyptians reverted to employing the 360-day year" (Warburton, 2012).

**The Khufu's granddaughter**

Let us consider Bauval, 2007. "Egyptologists and historians can never agree how old the Egyptian calendar is. There is, however, much evidence to support the conclusion that it was already in place during the Old Kingdom, for in the Pyramid texts there are several passages that allude to it indirectly … More direct evidence of the civil calendar in the Old Kingdom is found in the Fourth Dynasty tomb of Princess Mersyankh III [also Meresankh III] a [grand]daughter of King Khufu. An inscription on the entrance to her tomb at Giza, which was studied by the American Egyptologists Dows Dunham and William Kelly in 1974, given the date of her death … and the date of her burial" (Bauval, 2007). King's daughter Mersyankh, Year 1, month 1 of Shemu, day 21 (the death), Year 2, month 2 of Proyet, day 18 (the burial). The time between death and burial was of 273 days, that is, very close to nine months (Bauval, 2007).

In [Dunham and Simpson, 1974](#):

*North Side* (1) Vertical inscription, signs facing left:
S3t nswt Mr.sy-'nḫ ḥ3t-sp 1 3bd tpy šmw sw 21 ḥtp k3.s ḥpt.s r w'bt.
*Translation:* King's daughter Mersyankh, Year 1, month 1 of Shomu, day 21. The resting of her Ka and her proceeding to the house of purification (embalming).

*South Side* (2) Vertical inscription, signs facing right:
Ḥmt nswt Mr.sy-'nḫ ḥ3t (m)-ḫt sp tpy, 3bd 2-nw prt sw 18 ḥpt.s r is.s nfr.
*Translation:* King's wife Mersyankh. Year after 1 [Year 2], month 2 of Proyet, day 18. Her proceeding to her beautiful tomb.

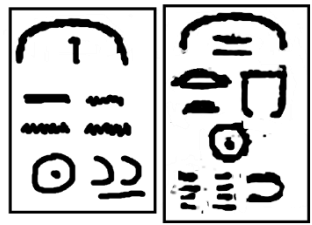

In Plate IIa of Dunham and Simpson, 1974, we can see the dates. On the left, the first month of Shemu, on the right second month of Peret (Proyet). Again, see Der Manuelian, 1986.

**Ne-User-Re (Neuserre, Niuserre)**

"Excavated by Egyptologists L. Borchardt and W. von Bissing beginning in 1898, the images and texts found at the sun temple of the pharaoh Ne-user-Re from the fifth dynasty in Abusir were a unique find. In 1974, E. Edel and St. Wenig published the first drawings and photos of the texts and images. Based on that edition, this book presents the first complete reconstruction and analysis of the extensive material found at the site" (Seyfried, 2019). This reference contains the Edel and Wenig work of 1974 (pictures from the original book are provided by [Meretseger Books)](#). The "head and ellipse" (as we have seen before), related to the epagomenal days, are visible in this plate. Wikimedia allows us to have a set of pictures of "Reliefs from Nyuserre Sun Temple in Berlin", Neues Museum.

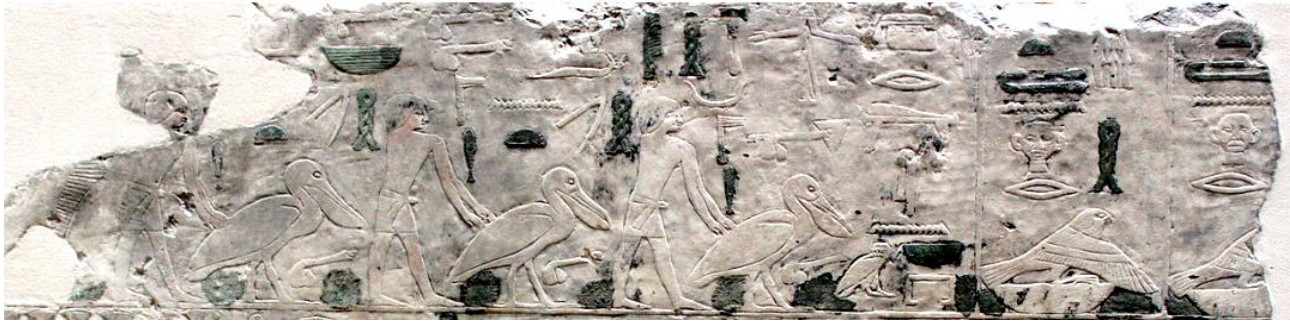

*Fig.12: The upper part of the "Fragment of a relief: breeding of pelicans and birds, fabrication of papyrus boats; 5th dynasty (ca. 2430 BC); Neues Museum Berlin, Germany; ÄM 20037, Courtesy Anagoria, CC BY 3.0.*

| | D2 U+13077 | face | face (ḥr) on, around, over, for (ḥr) | ḥr | 1. Bil. ḥr-(ḥr) 2. Ideogram for 'face' 3. A major preposition for "on, upon", etc.; additional preposition constructs. |
|---|---|---|---|---|---|
| | D21 U+1308B | mouth | mouth, utterance (r[ꜣ]) Turn about (pḫr) | r, jw (Ptolemaic) | Possibly ancestral to Proto-Sinaitic Pe and its descendants |
| | V28 U+1339B | a twisted wick | ḥ | | Uniliteral ḥ; eternity, or a long time period, (also variations of time periods, with tweaks of the seated man holding renpet-constructs) God Huh? Possibly ancestral to Proto-Sinaitic Heth and its descendants |

*List of Egyptian hieroglyphs – We have "hrjw" in the Fig.12.*

About the pelican, see Diab, 2017, in an article entitled "From an Immigrated Bird to a Deity: Pelican in Ancient Egyptian Sources". In Abdelhamid, 2016, we can see the personification of Seasons in the Ancient Egyptian Art. Note that in this article too, we can appreciate the seasons as represented in Neuserre Sun Temple. The name of the king is also written as Niuserre, Nyuserre, Niuserra. See also these pictures, courtesy kairoinfo4you. And Wikipedia about the Sonnenheiligtum des Niuserre. About Egyptian mythology and Neuserre see please Schott, 1945.

**Cattle Count**

Here are some considerations from Spalinger, 2018. "It would be useful if the early First and Second Dynasty records contained evidence for the use of the Civil Calendar. They do not. What we see instead is the gradual development of a supplementary bureaucratic system of calendrical notation: the regnal year. The first evidence for such notations occurs right at the beginning of the First Dynasty. We can ascertain from various ivory labels, vase inscriptions, and tables that each year of a king was given a specific name" (Spalinger, 2018). It is not possible to tell if the years were lunar or of a civil calendar. The Third Dinasty sees a consolidation of the state but "records are mute" regarding the calendar. Starting from the Fourth Dynasty, there is "civil reckoning chronicles in contemporary documents as well as in the later Palermo Stone. It is unknown whether this has to do with the rise of the cult of the sun, exemplified by pharaohs' relationship to the sun god Re, his father, and the sky goddess Nut, his mother. … At the end of the reign of Pepi II [Sixth Dynasty] it appears that the biennial system [of cattle census] was abandoned and that all regnal year dates referred to an individual year. The regnal year system enables the Egyptians to count each pharaoh's reign according to a regular system of calendrics; namely the Civil Calendar. In the Palermo Stone we find the scheme fully operational for the first time under the Fourth Dynasty ruler Shepseskaf; earlier entries are inscrutable" (Spalinger, 2018). Let us stress the Civil Calendar, that is 360-day year + five days, is also possible in the Cattle Count system. And in fact, in the Diary of Merer we find the thirteen cattle count in the reign of Khufu.

What Spalinger considers the Civil Calendar is the 360-day year + five days, with a number according to the reign of the king (for instance, Regnal year 1, fourth month of Inundation, day 5, under the majesty of King So-and-So). Spalinger does not mention Merer's Diary.

"In ancient Egypt, the cattle count was one of the two main means of evaluating the amount of taxes to be levied, the other one being the height of the annual inundation. A very important economic event, the cattle count was controlled by high officials and was connected to several cultic feasts. In addition, it served as a means of dating other events, with the entire year when it occurred being called "year of the Xth cattle count under the person of king Y". The frequency of cattle counts varied through the history of ancient Egypt; in the Old Kingdom it was most likely biennial, i.e. occurring every two years, and became more frequent subsequently" (Wikipedia and references therein).

The Cattle Count is important to evaluate the length of a reign. However, problems exist.

"An example of conflicting evaluations for a reign duration via cattle count is the case of king Khufu (4th Dynasty). The highest known numbers of cattle-counts under Khufu are found in workmen's graffiti inside the relieving chambers of the Khufu pyramid. The ink inscription reports the "17th occasion of the cattle count". Since the Palermo stone inscriptions hold that the cattle count was performed every second year during the 4th Dynasty, it would prove that Khufu ruled at least 34 years. This calculation is rejected by several Egyptologists, because another ancient Egyptian source, the Turin canon, credits Khufu with a reign of merely 23 years. …" (Wikipedia and reference therein).

**Civil versus Religious**

We can see from literature that, before the 365-day year, there was a simpler 360-day year, and consequently a calendar with twelve 30-day months. However, the Egyptians realized that this 365-day year "did not work either". As a consequence, "in economic, administrative and sometimes even astronomical texts, the Egyptians reverted to employing the 360-day year" (Warburton, 2012). Mesopotamia had an administrative calendar of 360 days too (Brack-Bernsen, 2007).

Now, let us consider literature about the coexistence of calendars in Egypt.

Stern, 2012. "A frequent model in the study of ancient calendars is that of the dual (or multiple) calendar, which posits the coexistence of two or more calendars in any given society, typically a 'civil' and a 'religious' calendar". The origin of this terminology seems to be in Censorinus (third century CE). "But the contrastive pairing of 'civil' with 'religious' is most likely the product of early modern scholarship, and as most ancient historians would now acknowledge, an anachronism". "More importantly", the hypothesis of concurrent calendars "is in many cases not grounded on any evidence, but only an awkward purpose-made solution to apparent inconsistencies in the primary sources". "The dual or multiple calendar model" "has been invoked by modern scholars in the context of just about every ancient society", that is, in Egypt, Mesopotamia, Israel and Judaea, Classical and Hellenistic Greece (Stern, 2012).

"Another well-established belief in ancient calendar studies is that modern astronomy can enable the date of institution of ancient calendars to be precisely worked out. Accordingly, the institution of the Egyptian civil calendar has been dated precisely to 2781–2778 BCE, when modern astronomy tells us that the Egyptian New Year coincided with the heliacal (late morning) rising of the star of Sothis" (Stern, 2012).

"Related to these astronomical arguments is what I [Stern] would call 'scientific reductionism', which consists in reducing ancient calendars (and hence their study) to a purely scientific discipline. Fostered by a nineteenth-century tradition of handbooks of so-called mathematische und technische Chronologie, [and Stern is mentioning in a note Ideler and Ginzel] this approach assumes that calendars were constructed and regulated on the sole basis of scientific, especially astronomical and mathematical, knowledge, and hence that their development was driven by scientific progress" (Stern, 2012).

"The inaccuracy of the Egyptian civil calendar as a solar calendar was evident to Egyptian astronomers and could easily have been corrected with the insertion of an additional day every four years, as in fact was proposed in the mid-third century BCE in the decree of Canopus; but" the Egyptian calendar remained "unchanged for several millennia, until the arrival of the Romans", and Romans acted "for political rather than astronomical reasons" (Stern, 2012).

"The Egyptian calendar was probably the simplest calendar of the ancient world. In contrast to the irregular, unpredictable calendars of Babylonia and Greece, it was regular, changeless, and fixed. Its structure was almost completely uniform: all months were of 30 days, and all years of 12 months. The only anomaly was the addition of 5 days (the 'epagomenal', i.e. additional, days) at the end of each year, so as to bring its total length to 365 days" (Stern, 2012). "The average or approximate solar year-length of 365 days is not very difficult, empirically, to work out, hence not much of an intellectual feat" (Stern, 2012). The Egyptian calendar has not an equivalent of the present leap years and therefore "this calendar drifted away from the solar year by one day every four years. The excessive praise that is commonly bestowed on this calendar, by Nilsson as well as other scholars, reflects to some extent a modern bias. The modern attraction to the Egyptian calendar is partly due to its structural and functional similarity to the present-day Gregorian calendar … This, however, is no historical accident, as the origins of the Gregorian calendar go back to the Egyptian" (Stern, 2012)

"The Egyptian calendar is so simple and non-problematic that it would appear, at first sight, not to demand any discussion or analysis. Yet for a number of reasons, it has been the focus of considerable debate among Egyptologists. Above all, there is the problem of the evidence. … The calendar can be identified in much earlier Egyptian texts, but the evidence is sporadic and usually open to more than one interpretation" (Stern, 2012).

***Today, we have the Merer's Diary and in it the Civil Calendar, without any doubt.***

"The civil calendar was extremely ancient, as evidence goes back to the third millennium BCE; but its origins are shrouded in mystery. On the one hand, the subdivision of its year into 30-day periods ('months') suggests that the calendar may have been originally lunar, … the addition of five epagomenal days would have represented an attempt to bring this calendar in line, albeit approximately, with the solar year. On the other hand, the Egyptian names of months suggest that this calendar may have been originally seasonal. … It is also possible that the calendar originally combined both features, i.e. lunar months, but with a system of intercalation that maintained these months in the same seasons. In the absence of evidence, the question remains open" (Stern, 2012). The names of the months are the ordinal numbers of them in the season, then the first month was I Akhet (Inundation Season).

"The entries in the royal annals of the Fourth and Fifth Dynasties (normally dated to the third quarter of the third millennium BCE) indicate, on the accession of a new king, that the year (from New Year to New Year) that he shared with his predecessor amounted to twelve months and five days. Although the length of these months is not specified, the addition of five days is characteristic of the civil calendar and cannot be explained as relating to any other calendrical system. More evidence from the Fifth Dynasty is the testament of the priest Nekonkh, where five epagomenal days appear to be positioned at the beginning of the year. A little later in the third millennium, inscriptions on Pepi II's pyramid (Sixth Dynasty) refer to the birth of the gods during the five epagomenal days, which again are specific to the civil calendar" (Stern, 2012).

"The *common scholarly view* is that the civil calendar was instituted even *earlier, around 2780 BCE*. This precise date is based on an astronomical argument", and Stern illustrates this argument that the researcher considered "unconvincing". See please the discussion in Stern, 2012. "Nevertheless, the traditional association of the New Year with the rising of Sothis suggests that the New Year of the civil calendar was meant to coincide with this astronomical event, and that it did so when the calendar was originally instituted" (Stern, 2012, remarking that Clagett is "rightly stressing that this theory remains conjectural").

According to Stern, "a more plausible theory of origins is that of Neugebauer". Neugebauer argued that the civil calendar could not have been established on the astronomical basis of the heliacal rising of Sothis, "because the calendar's drift of one day every four years would have disrupted the relationship between Sothis and the New Year almost from the outset. Instead, Neugebauer suggests that the calendar was originally based on a seasonal criterion, whereby the New Year, I Akhet, was intended to coincide with the beginning of the inundation of the Nile, as indeed the name Akhet ('inundation') suggests. Unlike the heliacal rising of Sothis, the inundation of the Nile *was* not a punctual event but a gradual process, which started every year in southern, upper Egypt a little after the summer solstice, and then spread downstream to lower Egypt during the following two months" (Stern, 2012). In Stern's note 26, it is stressed that, with the construction of the Aswan dams, "the inundation of the Nile has become completely disrupted". In this Stern's note, we find the author mentioning Belmonte Avilés (2003), because of a calculation related to the periodicity of inundation. "The beginning of inundation was not an event that could be precisely defined". If the intention was to "set the New Year at the beginning of inundation, it may have taken a long time for anyone to realize that the calendar was drifting away from it (by one day every four years), which means that the calendar would not have been invalidated from the outset. According to this theory, the institution of the calendar cannot be dated exactly to 2781–2778 BCE, but rather more loosely to some time in the early third millennium BCE, when I Akhet coincided roughly with the beginning of the inundation; any attempt at further precision is futile" (Stern, 2012).

**Wandering year**

"The year of 365 days was called the vague year, or the wandering year; and though the priests knew it to be inexact, they would not allow 'it to be altered by any system of leap years. At least there was a time in their history when this was the case. They required each Egyptian king,' on his accession to the throne, to bind himself by an oath, before the priest of Isis in the temple of Ptah at Memphis, not to intercalate either days or months, but to retain the year of 365 days as established by the Ancients (Mommsen, Chronologie, p. 258). Possibly their desire was simply to avoid confusion in the keeping of the feasts. Anxiety of this kind is manifested in the Decree of Canopus, on the Stone of San, where an annual festival is ordained in honour of Ptolemy Euergetes I and his queen, Berenice, and provision is made for an intercalary day to be added every fourth year, so as to keep the festival to the day of the rising of Sothis. But Mr. Lockyer thinks the priests generally were actuated by a love of power. They alone could tell on what particular day of what particular month the Nile would rise in each year, because they alone knew in what part of the cycle of 1,460 years they were ; and in order to get that knowledge they had simply to continue going every year into their Holy of Holies one day in the year, and watch the little patch of bright sunlight coming into the sanctuary. That would tell them exactly the relation of the true summer solstice to their year, which was supposed to begin at the solstice; and the exact date of the inundation of the Nile could be found by those who could determine observation- ally the solstice, but by no others. It would appear, however, that the period of six hours was in some way recognized, and the existence of the hours set forth symbolically; for Mr. Hilton Price describes in his collection "an emblem of Mahen, goddess of the hours, wearing the disk and cow-horns ; and upon the breast are six compartments, which have been filled up with glass and stone."" (St Clair, 1898).

**2773 BC**

Let us return to the article of Darvill about Stonehenge. "Back-calculations using the pre-existing lunar-stellar calendar suggest that the Civil Calendar started in 2773 BC. Like the cult of Ra, with which it was closely associated, the Civil Calendar was widely used during the Third Dynasty at the start of the Old Kingdom, c. 2658 BC. By the Fifth Dynasty (c. 2500–2300 BC), the cult of Ra had become the state religion, with rulers taking the title 'Son of Ra' (Quirke, 2001: 17)" (Darvill, 2022). About the year 2773 BC, it is evident that punctual information is lacking in this part of Darvill's text. But Darvill is mentioning Stern, 2012, in relation to the Civil Calendar (see please above). At the same time, reviewer and editor (Antiquity Journal) have not asked the author, Darvill, to provide further specific detail. May be, it was just a problem coming from a limited length of manuscript. In any case, the year 2773 BC is highly intriguing, being not so far from Khufu's time. Let us add further information, besides Stern's passages given above.

"In 1904 Eduard Meyer stated that the Egyptian calendar was invented about 4231 BC, and some of the principal Egyptologists of his generation adopted this theory with minor modifications. In recent years it has been realized that 4231 BC was far back in the prehistoric period, long before the invention of writing, and of necessity later dates have had to be advanced for the adoption of the calendar as we know it. Primitive man in Egypt regulated his life entirely by the cycle of the Nile's stages. Nature divided its year into three well-defined seasons-Flood, Spring, and Low Water or Harvest, with the Flood Season, following the hardship of the Low Nile, the obvious starting point for each annual cycle. The Egyptian early recognized the fact that usually twelve moons would complete a Nile year, but his lunar reckoning always remained secondary to his Nile reckoning, and he never adopted solar seasons. However, by about 3200 BC he probably recognized the heliacal rising of the prominent star Sothis as a definite phenomenon heralding the coming flood, and he began to count the observed reappearance of the star as his New Year Day. His year he now adjusted to twelve artificial moons of 30 days each, followed by about five days in which he awaited the reappearance of Sothis [Sirius star]. For several centuries the calendar was fixed to the star and thus was approximately correct, but the experience of generations was apparently proving that the perfect year should be 365 days long, and in 2773 BC a year of this length was adopted, by the simple expedient of neglecting to readjust the calendar by annual observations. Since no change was ever permitted thereafter, the Egyptian calendar was only correct once in every 1460 years." (Abstract of the article by Winlock, 1940).

Year 2773 BC is regarding therefore the [Sothic Cycle](). Adapting from Wikipedia (see please references therein): The Sothic Cycle was "well known in antiquity. Censorinus described it in his book De Die Natale", as we have already seen in Stern, 2012. "In the ninth century, Syncellus epitomized the Sothic Cycle in the Old Egyptian Chronicle. Isaac Cullimore, … Royal Society, published a discourse on it in 1833 in which he was the first to … [He] computed the likely date of its invention as being around 1600 BCE. In 1904, seven decades after Cullimore, Eduard Meyer carefully combed known Egyptian inscriptions and written materials to find any mention of the calendar dates when Sirius rose at dawn. He found six of them, on which the dates of much of conventional Egyptian chronology are based. … Meyer concluded that the Egyptian civil calendar was created in 4241 BCE. Recent scholarship, however, has discredited that claim. Most scholars either move the observation upon which he based this forward by one cycle of Sirius, to 19 July 2781 BCE, or reject the assumption that the document on which Meyer relied indicates a rise of Sirius at all" (Wikipedia and references therein).

"Three specific observations of the heliacal rise of Sirius are extremely important for Egyptian chronology. The first is the ivory tablet from the reign of Djer which supposedly indicates the beginning of a Sothic cycle, the rising of Sirius on the same day as the new year. If this does indicate the beginning of a Sothic cycle, it must date to about 17 July 2773 BCE" (Wikipedia; Grimal, 1988). "However, this date is too late for Djer's reign, so many scholars believe that it indicates a correlation between the rising of Sirius and the Egyptian lunar calendar, instead of the solar Egyptian civil calendar, which would render the tablet essentially devoid of chronological value" (Wikipedia; Grimal, 1988).

Since we have mentioned the lunar calendar, let us consider Depuydt, 1997, Civil calendar and Lunar calendar in Ancient Egypt. "Before writing, in the fifth and fourth millennia BCE, the calendar of daily life in Egypt was the lunar calendar in one or more of its possible variations. Sometimes early in the third millennium, writing was invented. Also invented around the same time was a calendar that was independent from the moon. It was inspired by the lunar calendar as the lengths of its months show. The new calendar consisted of 12 months of 30 days and 5 epagomenal days for a total of 365 days. This simple calendar, now known as the civil calendar, became the calendar of daily life for the entire course of the Egyptian history. Being 365 days long, it was designed to run parallel to the solar year. But because it was about a quarter day shorter, it wandered in relation to the solar year" (Depuydt, 1997).

"At the same time, in spite of the dominance of the civil calendar, there was always lunar time-reckoning. But the lunar calendar, once the popular calendar of Egypt in the prehistorical period, was now reduced to the role of a religious calendar and restricted mainly to life in the temple" (Depuydt, 1997).

In DeYoung, 2000, regarding the Sothic Cycle, it is said that the lunar calendar "was not suited for administrative use". Then the Egyptian bureaucracy adopted a calendar made of 12 months of 30 days, plus five epagomenal days. We have 365 days in total. "Why 365 days? Richard Parker suggests (1971) the figure might have arisen either from an averaging of the number of days in several successive luni-stellar years or from counting the number of days between successive heliacal rising of Sothis. Since this civil calendar will lose approximately a quarter of a day each sideral year, it will require a period of 1461 administrative years (the so-called Sothic cycle) until the two calendars fall into step once again. We know from historical sources that in 139 CE the beginning of the civil calendar year coincided with the heliacal rising of Sothis. From this fixed reference point, we can calculate that these two events occurred together also in 1317, in 2773, and in 4229 BCE (Taton, 1957, 1963). Since our earliest literary source mentioning a correlation between the two calendars dates from the reign of Sesostris (or Senusret) III (1874-1855 BCE), the civil calendar must have been introduced prior to that time. Parker suggests (1971) that the civil calendar may have been initiated a century or two prior to 2800 BCE rather than at the earlier date. Clagett, on the other hand, believes (1995) that the civil calendar may date back to the pre-dynastic era (4000-3000 BC)" (DeYoung, 2000).

**Today, we have the Merer's Diary and in it the Civil Calendar, without any doubt.**

Let us add that the Sothic Cycle has been illustrated in his Nobel Lecture by Ahmed H. Zewail, entitled "Femtochemistry: atomic-scale dynamics of the chemical bond using ultrafast lasers". Please, consider a careful reading of Zewail's Nobel Lecture. "Ahmed H. Zewail (born February 26, 1946, Damanhur, Egypt—died August 2, 2016, Pasadena, California, U.S.) was an Egyptian-born chemist who won the Nobel Prize for Chemistry in 1999 for developing a rapid laser technique that enabled scientists to study the action of atoms during chemical reactions. The breakthrough created a new field of physical chemistry known as femtochemistry. Zewail was the first Egyptian and the first Arab to win a Nobel Prize in a science category" (Britannica).

Read please the description of the Egyptian civil calendar in the Nobel Lecture by Zewail. Here just the beginning: "My ancestors contributed to the beginning of the science of time, developing what Neugebauer had described as "the only intelligent calendar which ever existed in human history".

"The 'Nile Calendar' was an essential part of the life as it defined the state of yearly flooding with three seasons, the Inundation or Flooding, Planting, and Harvesting, each four months long. A civil year lasting 365 days was ascertained by about 3000 BC or before, based on the average time between arrivals of the flood at Heliopolis, just north of Cairo; nilometers were used in more recent times, and some are still in existence today. By the time of the first Dynasty of United Egypt under Menes is about 3100 BC, the scientists of the land introduced the concept of the "Astronomical Calendar" by observing the event of the heliacal rising of the brilliant star Sothis (or Sirius). Inscribed on the Ivory Tablet, dating from the First Dynasty and now at the University Museum in Philadelphia, were the words, "Sothis Bringer of the Year and of the Inundation. … Thus, as early as 3100 BC, they recognized a definite natural phenomenon for the accurate timing of the coming flood and recounted the observed reappearance of the star as the New Year" (Zewail, Nobel Lecture 1999, mentioning Winlock).

**Heliacal rising on a Jar**

Gautschy et al., 2017. "A recently discovered inscription on an ancient Egyptian ointment jar mentions the *heliacal rising of Sirius*. In the time of the early Pharaohs, this specific astronomical event marked the beginning of the Egyptian New Year and originally the annual return of the Nile flood, making it of great ritual importance. Since the *Egyptian civil calendar* of 365 days permanently shifted one day in four years in comparison to the stars *due to the lack of intercalation*, the connection of a date from the Egyptian civil calendar with the heliacal rising of Sothis is vitally important for the reconstruction of chronology. The new Sothis date from the Old Kingdom (3rd–6th Dynasties) in combination with other astronomical data and radiocarbon dating re-calibrates the chronology of ancient Egypt and consequently the dating of the Pyramids. A chronological model for Dynasties 3 to 6 constructed on the basis of calculated astronomical data and contemporaneously documented year dates of Pharaohs is presented" (Gautschy et al., 2017). In the abstract, Gautschy and coworkers mention the civil calendar but, in the article, regarding this calendar, they do not provide any reference about the lack of intercalation. They discuss the lunar calendar and the intercalation. We are induced to assume, without evidence, that lunar and civil calendars coexisted.

The calendar of Wadi el-Jarf calendar was the civil calendar (Tallet, 2017, about Papyrus H).

"Recently a new Sothis date on an ointment jar was discovered [Gautschy and coworkers mentioning Habicht, et al., 2015]. The jar mentions the "Forthcoming of Sopdet" and the date of a heliacal rising of Sirius on the beaker. For the stylistic dating of our jar several publications were used [see references in Gautschy et al.]. Such cylindrical beakers contained perfume oil and were often given to people on special occasions and festivals—a tradition still alive in the Coptic church. The jar was stylistically dated into the mid to late 5th Dynasty. Inscription, palaeography and the astronomical date also point to the Old Kingdom" (Gautschy et al. and references therein).

Gautschy and coworkers used documents "from Dynasties 4 to 6", that are "mentioning about 150 different years of Pharaohs". The researcher used compilations prepared by Spalinger, Verner, and Gundacker (Gautschy et al. and references therein). In their Table II, the authors provide a "list for all years between 2636 and 2280 BCE with the Egyptian date of the beginning of the *lunar month in which the heliacal rising of Sirius presumably occurred*, the dates of the heliacal rising of Sirius for an arc of vision of 9° to 10°, the number of lunar months in an Egyptian year, the year count of pharaohs Khufu to Pepy II according to *our [Gautschy and coworkers] High Chronology* as well as the documented regnal years, their source, and remarks (uncertain attributions and astronomical data)" (Gautschy et al., 2017). In that Table, we find the first mention of Khufu associated to year 2636.

Let us repeat a passage from Robinson, 2022. "Without a doubt, the Great Pyramid was commissioned by the Old Kingdom pharaoh Khufu (Cheops). The British Museum and Cairo's Egyptian Museum give his regnal dates as 2589 to 2566 BCE. Egyptologists Mark Lehner, who has conducted fieldwork at Giza for four decades, and Zahi Hawass, a former Egyptian government official in charge of Giza, argued for the later range of 2509 to 2483 BCE in their massive 2017 book, Giza and the Pyramids. But another high-profile Egyptologist, Pierre Tallet, whose pioneering fieldwork on the Red Sea coast of Egypt began in 2011, favors the earlier range of 2633 to 2605 BCE, derived from a recent astronomically based chronological model for the Old Kingdom" (Robinson, 2022). It seems that the model of astronomical chronology is based on the heliacal rising of Sirus then,

Let us stress once more that the calendar that we find in the Wadi el-Jarf papyri is the Egyptian Civil Calendar (see Tallet, 2017, about Papyrus H).

In Table 3 by Gautschy et al., 2017, regarding the "Absolute dates and number of counted regnal years in our [Gautschy et al.] Low Chronology (left) and our High Chronology (right) of the Old Kingdom", we find Khufu according to the following screenshot.

| | Low chronology | Hornung et al. | Shaw | von Beckerath | Dee (C14: 95%) | High chronology |
|---|---|---|---|---|---|---|
| Khufu | 2503 BCE | 2509 BCE | 2589 BCE | 2604/2554 BCE | 2629–2558 BCE | 2636 BCE |

*This is the upper part of Table 4 in Gautschy et al. (see please references therein).*

**Egyptologists and Anthropologists**

The proposal by Darvill of a Stonehenge solar calendar received some observations made by two archaeoastronomers, Belmonte and Magli, (arXiv, 2022) that Darvill answered in 2023. Darvill notes that, "Supported mainly by self-citation, the authors [Belmonte and Magli] claim that the basic calendar of twelve 30-day months (each comprising three 10-day weeks) together with five epagomenal days, and perhaps one additional day every four years, is a scheme not represented until the later first millennium BC. This view is at odds with most other authorities on the subject" (Darvill, 2023). *We have the Merer's Diary and in it the Civil Calendar, without any doubt. And the time was that of Akhet Khufu's building. Note that in the Merer's Diary we find 30-day months, three 10-day weeks in a month, and epagomenals. Season and month names, such as Epagomenals, are according to inscriptions on the east wall of the Tehne tomb of Nj-k'-'nh, early Fifth Dynasty.*

Let us consider other observations by archaeoastronomers. In Darvill, 2022, we read: "Twelve monthly cycles of 30 days, represented by the uprights of the Sarsen Circle [Stonehenge], gives 360 solar days. While no stones within the central setting can specifically be identified with the 12 months, it is possible that the poorly known stone settings in and around the north-eastern entrance somehow marked this cycle" (Darvill, 2022, and references therein). Archaeoastronomers Belmonte and Magli observe that "the number 12 is entirely absent from the [Stonehenge] monument". Darvill has already informed us that number 12 was absent. *Number 12 was not necessary to be explicitly stated in the Egyptian calendar, where the months had numbers from the seasons (I-Akhet, II-Akhet, III-Akhet,*

*IV-Akhet, and the same for the other two seasons, Peret, and Shemu). As previously told, Stern, 2012, mentions Belmonte, 2003, and Belmonte is giving three seasons, each with four months.*

One observation is regarding the year 2773 BC that we have discussed above: "The starting date of the civil calendar, for example, is much debated and it would be far-fetched to fix such a precise date as 2773 BC". Belmonte and Magli, in formulating this observation, do not provide any information but tell "No Egyptologist would accept such an assertion" (self-citations). Again, reviewer and editor (Antiquity Journal) did not asked Belmonte and Magli to be more specific about year 2773 BC, turning the "No Egyptologist …" sentence in an authority fallacy (that is, the logical fallacy of saying that a claim is true simply because an authority figure, in this case the Egyptologist, is inserted in it).

*Can't we set a specific year? Surely Merer had a calendar with 30-day-months, 10-day-weeks, epagomenals, seasons and month names according to inscriptions on the east wall of the Tehne tomb of Nj-k'-'nh, early Fifth Dynasty. The timekeeping at Akhet Khufu was based on the civil calendar (that of workers at the pyramid). This is what we find in Tallet, 2017, about Papyrus H.*

There is another passage in the observations by Belmonte and Magli where reviewer and editor (Antiquity Journal) had not asked for more detail. "Such a transfer and elaboration of the Egyptian calendar c. 2600 BC resembles old ideas of diffusionism that *we [Belmonte and Magli]* believed had been discarded and happily forgotten" (no references given).

Belmonte and Magli take diffusionism from the very Darvill's text of 2022, where we read: "Archaeologically, the question is whether the Egyptian Civil Calendar, or a variation thereof, could have been known to communities living in southern Britain in the mid-third millennium BC, and adopted by them. Barely a century ago, the answer would have been resoundingly affirmative (e.g. Childe, 1929). As diffusionist models crumbled and connections between the Mediterranean world and Northern Europe were systematically uncoupled to emphasise autonomous local development (Renfrew,1973: 84–108), such thinking became deeply unfashionable. Now, however, the pendulum of interpretation is swinging back in favour of long-distance contacts and extensive social networks. Tight radiocarbon chronologies allow synchronicities to be recognised, while aDNA and isotope studies emphasise the nature and extent of population movements during the third millennium BC" (see Darvill, 202, and references therein).

Diffusionism means Anthropology. Let us then report the abstract by Rohatynskyj, 2018, for illustrating the subject. "Diffusionism arose in the formulation of the discipline of anthropology as an explanation of cultural similarity across geographical regions. It posited that elements of culture, often termed "culture traits," were invented once and spread out to neighboring groups undergoing adaptation in their progress. Early diffusionists are referred to as "extreme diffusionists" in the present because their accounts of historical movements of traits around the globe were simplistic and largely unprovable. Diffusionism, in the present, serves as a yet-to-be-disproven assumption about the nature of human interaction across cultural boundaries in that the tendency to culture borrowing and sharing appears a universal. It also serves as the basis of a methodology which underlies all four subfields in anthropology, as well as geography and history, in tracing cultural elements geographically and through time. The diffusionist premise of the historical unity of humankind remains critical in the contemporary globalized world" (Rohatynskyj, 2018).

And Egypt? The web site [SociologyGuide](), 31 October 2024, tells: "Diffusionism refers to the diffusion or transmission of cultural characteristics or traits from the common society to all other societies. They [the diffusionists] criticized the [Psychic unity]() of mankind of evolutionists. They believed that most inventions happened just once and men being capable of imitation, these inventions were then diffused to other places. According to them all cultures originated at one point and then spread throughout the world. … The main proponents of British school of Diffusionism were G. Elliot Smith, William J. Perry and W.H.R. Rivers. They held the view that all cultures originated only in one part of the world. Egypt was the culture centre of the world and the cradle of civilization. Hence human culture originated in Egypt and then spread throughout the world. They pointed to the Pyramid-like large stone structures and sun worship in several parts of the world".

The SociologyGuide continues: "The German School of Diffusionism has chief proponents like Friedrich Ratzel, Leo Frobenius, Fritz Graebner and William Schmidt. Their approach was through the analysis of culture complexes identified geographically and studied as they spread and developed historically. It has both time and space dimensions. The first dimension of space was explained in terms of culture circles and the second dimension of time was explained in terms of culture strata". Let us just remember Frobenius.

In 1913, "The voice of Africa" by Leo Frobenius has been published. It was an account of the travels of the German Inner African Exploration Expedition in the years 1910-1912, made under Frobenius' leadership. Frobenius carried out excavations at the ancient Yoruba city of Ife in Nigeria, theorizing that the Ife bronze and terracotta sculptures were relics from Atlantis. In Sparavigna, 2024, we have stressed Frobenius knew the theory by Heinrich Nissen of the Templum, published in 1869. He used Nissen's templum as evidence of an Atlantic Africa connection with the Mediterranean world.

In his book then, Frobenius proposed the *diffusion* of the idea of the templum, as proposed by Nissen in 1869, from the Mediterranean Sea to Atlantic Nigeria.

Of Heinrich Nissen, Clive Ruggles says that he "deserves more than anyone else to be recognized as the earliest pioneer of modern archaeoastronomy". In fact, the German professor of ancient history studied the orientations of Egyptian and Greek temples. However, Nissen is rarely mentioned in accounts of the early development of ideas and methods for archaeoastronomy. For instance, until my works of 2020, it was **totally** ignored by archaeoastronomers what Nissen, in his Das Templum, 1869, proposed about the direction of the decumani of Roman towns (Sparavigna, 2020, 2024). Nissen was so ignored that, in [2007](#), Magli proposed the same theory without mentioning Nissen. Unlike today archaeoastronomers, Friedrich Nietzsche knew very well the Das Templum and Nissen's theory about the decumanus sunrise orientation on a day of a calendar festival. After the publication of the Das Templum, Nissen's theory was criticized, with well-posed arguments, by Valeton, 1893. For a specific discussion about templum, decumanus, pomerium, augural law, inauguration and dedication, see please Catalano, 1978.

The case of Nissen and the archaeoastronomers shows that the bibliographical references proposed in archaeoastronomical works must be verified with extreme attention, and one must also go further and expand the bibliography.

**The five-seasons calendar**

Today, seasons are four, but in ancient Egypt seasons were three. Why not five?

"The cultures in the flood plains of the *Nile and the Niger Rivers* in Africa have a natural three-season year: flood season, planting season, and harvest-fallow seasons. The Dog Star marks the inundation of the flood coming down the Nile and Niger Rivers from the tropical summer. This was called the season of Djehuti. It was also referenced to the ass-eared Set, because the dry winds of the high summer attract violent attacks from the denizens of the desert, the original Semites. The self-same weather conditions that create the rains in the tropics and flood the Nile also bring the jackals out of the desert. The Egyptian three-season calendar would have four months of 30 days each for every season, or twelve months all together. This gave 360 days with five left over. The Egyptian said that the five extra days in the calendar were those that Thoth won at draughts from the African moon-goddess Isis thereby honoring the calendar's northern imported origin. … These five days became the birthdays of Osiris, Horus, Set, Isis and Nephthys. This still left an error of closure of about a quarter of a day. When or how this error was corrected is unknown" (Becker, 2004).

Becker, 2004, is also referring to Graves' White Goddess. Robert Graves proposed "the existence of a European deity, the "White Goddess of Birth, Love and Death", much similar to the Mother Goddess, inspired and represented by the phases of the Moon, who lies behind the faces of the diverse goddesses of various European and pagan mythologies" ([Wikipedia.org](Wikipedia.org)).

"When Robert Graves wrote 'The White Goddess' he was under the influence of academy that confidently supposed that all progress in European intellectual and spiritual development came from the Near East. Graves supposed that the 360-day-plus-five calendar came to the north from the Egyptians. As we now know *Stonehenge was contemporaneous with the beginning of Sumer and Egypt, not a poor imitation at a later date*. As we now know from Gosekhenge is many centuries earlier that Sumer and Egypt. Placing the original pole star in Lyra gives the provenance of the first Egyptian calendar to the Anatolian cattle civilization around 7000 BC. Until it is positively demonstrated to be otherwise, we must assume that the civilization impetus came from the north to the Nile. This actually suits the creation stories that Sumer and Egypt tell about themselves, that they were civilized from the outside in" (Becker, 2004).

In Becker's book we can find also the Great Pyramid. "The main viewing gallery in the Great Pyramid was aimed at the Pole Star that was then in the constellation Draco. The Queen's Gallery was aimed at the Dog Star". Then, we find mentioned that the Egyptian three seasons were "in a dramatic contrast to the mother of all calendars, the ancient, *five-season* calendar of the north".

Five-seasons? "Robert Graves deconstructed the secrets of the Ninefold Muse of the Bronze Age in the White Goddess. The magical formulas were bound up with the tree calendar that reflected the New Forest in Europe. The Beth-Luis-Nion tree calendar had a year of 364 days plus one day for the correction and sacrifice in the winter. This calendar had five seasons. The Boibel-Loth tree calendar had a year of 360 days plus five days for correction and sacrifice in the winter. This calendar also has five seasons. We have traced the five-season year back to the Magdalenian Age. The Blanchard bone marked the number of lunations in the season of a five-season year – approximately seventy-two. There is no reason to assume that the number was any different at Goseckhenge, *Stonehenge*, built many centuries later, had an inner horseshoe of *five great trilithons* referencing to the canonical number of 72 days in a season. Even in historical times the Norse still used the five-season, ten-month calendar year. Each month comprises six weeks. Each week was five days: Tysdag, Odinsdag, Thorsdag, Frjadag and Thvarrdag or washing day when the deities rested. This accounted for 360 days. The extra whole days were accounted for by a three-day Yule sacrifice called Midsvetrarblot, a spring sacrifice about the middle of April called Sigrblot, and a late harvest sacrifice in October called Vertarblot. How the remainder of a quarter of a day was made up in not known" (Becker, 2004).

I have reported some passages from the book by Becker, 2004. Of course, it is necessary to check all what we find in it. Let us note a passage: "*Until it is positively demonstrated to be otherwise, we must*

*assume that the civilization impetus came from the north to the Nile*". The burden of proof (onus probandi incumbit ei qui dicit, non ei qui negat) lies with the one who speaks, in this case J. Becker, not the one who denies. That is, it is the obligation on a party in a dispute to provide sufficient warrant for its position (see more in [wikipedia](#)).

In the introduction we have mentioned the Göbekli Tepe calendar. The burden of proof is with the persons proposing the calendar. And the calendar is not the only problem regarding the interpretation of the site. Please consider reading the article entitled "Paradise Found or Common-Sense Lost? Göbekli Tepe's Last Decade as a Pre-Farming Cult Centre", by Edward B. Banning. In particular, please consider carefully the part of the article regarding archaeoastronomy.

**Academy and freedom**

Darvill, 2023: "In their conclusions, Magli and Belmonte propose that "matters such as ancient calendars, astronomical alignments and cultural astronomy should be reserved for specialists" in a way that suggests academic arrogance. Their attempts to undermine the central idea of a Stonehenge calendar … Most notable of all, however, is that Magli and Belmonte do not make any suggestions as to what the settings at Stonehenge might have meant, how they might have worked, or how they might have been used by prehistoric communities". Darvill concluded questioning whether archaeo-astronomy can do better. "Surely modern archaeoastronomy can do better?"

Here Belmonte and Magli words: "*We believe that* matters such as ancient calendars, astronomical alignments and cultural astronomy should be reserved to specialists, *trained in the subject, and not left to researchers from other disciplines, however renowned and knowledgeable in their own fields*".

What Magli and Belmonte believe is not relevant. In [Britannica](#), we find a definition of "Academic freedom". It is "the freedom of teachers and students to teach, study, and pursue knowledge and research without unreasonable interference or restriction from law, institutional regulations, *or public pressure*. Its basic elements include the freedom of teachers to inquire into any subject that evokes their intellectual concern; to present their findings to their students, colleagues, and others; to publish their data and conclusions without control or censorship; and to teach in the manner they consider professionally appropriate. For students, the basic elements include the freedom to study subjects that concern them and to form conclusions for themselves and express their opinions".

We all (not only academy) are free to inquire into any subject that evokes an intellectual concern. Darvill investigated the possibility of a calendar embedded in the sarsen structures of Stonehenge, that could be contemporary of the civil Egyptian calendar. Belmonte and Magli feel free to criticize Darvill, even in a harsh manner indeed. And Darvill free to respond in proper and polite manner. Martin Sweatman, a theoretical physicist, is free to propose his Göbekli Tepe calendar. In my case, the main inquiry concerns the Roman templum; in this framework I investigate how Nissen's Das Templum influenced cultural approaches (as in the case of Leo Frobenius). For instance, an often-mentioned article by archaeoastronomers is that written by Aveni and Romano, 1994, where we can find Nissen and his Das Templum, 1869, among the given references. Aveni and Romano are referring just to the data provided by Nissen's book about the orientation of temples, and nothing more, of the large work made by Nissen.

In Aveni and Romano we find a proposal which is in Nissen's Das Templum, 1869, but it is not told that Nissen had already discussed it. "There is yet another astronomical possibility [for the orientation of the temples]. The most easterly temples (and those close to west for that matter) could be oriented to the sunrise (sunset) not on the equinox, but on dates that correspond to certain sacred festivals, perhaps left over from an archaic solar cult" (Aveni and Romano, 1994). A reader could imagine that the proposal by Aveni and Romano, regarding the link between temples and festival, was dated 1994, but it is not so. It is dated 1869, Das Templum, Nissen (detailed discussion in Sparavigna, 2024, about the 'liver of Piacenza'). The origin is Nissen. The Das Templum had been criticized. It is fundamental to know that criticism exists. It is essential to read the criticism (for a detailed discussion of the Roman Templum see please Catalano, 1978, and my discussion of literature in "The Town and the Templum", 2024).

The observation about Aveni and Romano, archaeoastronomers, is evidencing that it is basic to check the references given in an article and to go beyond them, to become a person *trained on the subject*, and this is true for the Roman world, for the ancient Egyptian world, for ancient Stonehenge, for Göbekli Tepe, and so on. Moreover, and in a general case, it is fundamental to recognize the fallacies used in argumentation. Being trained in identifying fallacies allows us to develop a "critical thinking", so to be able not to take everything as gospel, even if it is published in a peer-reviewed article.

**Conclusion**

Here we have shown that the Egyptian Civil Calendar is present in the Merer's diary, discovered and studied by Pierre Tallet (2017). Then, let us repeat what Darvill noted in 2023: "Supported mainly by self-citation, [archaeoastronomers Belmonte and Magli] claim that the basic calendar of twelve 30-day months (each comprising three 10-day weeks) together with five epagomenal days, and *perhaps* one additional day every four years, *is a scheme not represented until the later first millennium BC*. This view is at odds with most other authorities on the subject" (Darvill, 2023). At the time of Khufu, that is, circa 2600 BC, we can tell that there were 30-day-months, 10-days weeks, epagomenal days, and that the yearly missions of Merer and his crew had been organized according to the seasons. The months are those that we find in the Nj-k-'nh's table (early Fifth Dynasty), and also Epagomenals. Number 12 was not disclosed in the calendar, because seasons were fundamental, in the number of three, and each season has four months. Other evidence, as given by literature, indicates the existence of five epagomenal days at the end of the fourth or the beginning of the fifth dynasty.

Please, do not confuse the Egyptian civil calendar with the Julian Calendar. In Darvill, 2022, the Julian calendar is not mentioned. Archaeoastronomers Belmonte and Magli mention the Julian calendar, saying that the "first documented elaboration of such a calendar dates to two millennia later, when an unsuccessful attempt to implement it was made in Ptolemaic Egypt, with the so-called Canopus Decree". Please note that Canopus Decree is regarding the Egyptian Civil calendar (wandering calendar). This calendar existed quite before the Canopus decree. And Merer, at Khufu's time, used 30-day-months, 10-day-weeks, five epagomenal days, three 4-month-seasons [1].

---

[1] To the best of my knowledge, Belmonte and Magli did not consider Tallet's discovery for what is regarding the calendar, but they just mention the papyrus about "Akhet Khufu". In 2015, in note of an article entitled "Astronomy, architecture, and symbolism: the global project of Sneferu at Dahshur", Belmonte and Magli are reproposing the idea that the two largest pyramids at Giza were planned in a single project by Khufu. Referring to "The harbour of Khufu on the Red Sea coast at Wadi al-Jarf, Egypt", Near Eastern Archaeology, the two archaeoastronomers tell that Tallet and Marouard "have found a papyrus, contemporaneous to Cheops's reign, with a text where Axt xfw is mentioned with a single determinative of a pyramid (as in later sources). This,

For what is regarding the wandering of Egyptian Civil Calendar, it is possible to argue that adjustments and changes were often made, "usually at critical moments such as the start or end of a year, in order to keep everything running smoothly" (Darvill, 2023). An example of running smoothly a calendar is Augustus' reform of the Julian Calendar (Polverini, 2016). Here it is what Leandro Polverini says, translated from Italian. One of the most curious facts in the history of calendars - in general, full of curious facts - and of the Roman calendar in particular, is that the pontiffs in charge of running the calendar inserted every three years, instead of every four years, the intercalary day prescribed by the reform. The error lasted for 36 years, from 45 to 9 BC, during which twelve triennial intercalations were made, instead of nine four-year intercalations. In 36 years, the official calendar had a lag of three days. In 8 BC Augustus intervened in his own way: instead of subtracting three days from the calendar (Caesar had not hesitated to add 90 days in a single year!), he decided to suspend the intercalation three times. After the intercalation of 9 BC, those of 5 BC, 1 BC, and 4 AD were not made. With the intercalation of 8 AD, the Julian calendar – more than fifty years after Caesar stated it – was finally fully operational. Then, even the Julian calendar required adjustments; in the same manner, also the Egyptian Civil Calendar could had been run with smooth variations.

Let us return to Stonehenge. "The smaller "bluestones" near the center of the monument have been traced to Wales, but the origins of the sarsen (silcrete) megaliths that form the primary architecture of Stonehenge remain unknown" (Nash et al., 2020). Nash and coworkers "use geochemical data to show that 50 of the 52 sarsens at the monument share a consistent chemistry and, by inference, originated from a common source area. [Nash and coworkers] then compare the geochemical signature of a core extracted from Stone 58 at Stonehenge with equivalent data for sarsens from across southern Britain. From this, [Nash and coworkers] identify West Woods, Wiltshire, 25 km north of Stonehenge, as the most probable source area for the majority of sarsens at the monument" (Nash et al., 2020). We have not written text, but a planning supervisor existed, to organize the crews and their mission. What was the best season for their task? Stonehenge Sarsen circle had been planned with a geometry based on number 30, and to my best knowledge no one raised any concern regarding this planning. We must be impressed by the layout (see the [figure](#) in Darvill, 2022), so that it is possible to argue that people at Stonehenge developed geometry and numerals, and a calendar too. People at Göbekli Tepe were able of using a solar calendar, so could at Stonehenge.

*Verba volant, scripta manent*

**References**

1. Abdelhamid, T. (2016). Personification of Seasons in the Ancient Egyptian Art. Journal of

---

according to José Lull (private communication), would be a point against the duality of the complex". Please consider again the article by Lehner about Merer and the Sphynx. In a book by Belmonte and Lull, 2023, "Astronomy of Ancient Egypt", we can find in the Introduction, mentioned "Taller" and Lehner, 2021. Belmonte and Lull are just telling that surprises are always possible as "demonstrated by the recent discovery of papyrus fragments of Khufu's pyramid accountancy in a barren spot of the Red Sea coast (Taller & Lehner, 2021)". Nothing more. The "barren spot" is Wadi al-Jarf that hosted the artificial harbor of Khufu, where phyles linked Egypt to Sinai.